\documentclass[apj,numberedappendix,appendixfloats,twocolappendix]{emulateapj}

\usepackage{amsmath}

\newcommand {\eV}       {\,\rm eV}
\newcommand {\pc}       {\,\rm pc}
\newcommand {\kpc}      {\,\rm kpc}
\newcommand {\Mpc}      {\,\rm Mpc}
\newcommand {\kpch}     {\,h^{-1}\,\rm kpc}
\newcommand {\Mpch}     {\,h^{-1}\,\rm Mpc}
\newcommand {\Msun}     {\, M_{\sun}}
\newcommand {\kms}      {\,\rm km\,s^{-1}}

\newcommand {\psiDM}    {\psi {\rm DM}}
\newcommand {\PM}       {m_{\rm 22}}
\newcommand {\Mlim}     {M_{\rm lim}}
\newcommand {\Llim}     {L_{\rm lim}}
\newcommand {\QHII}     {Q_{\rm HII}}
\newcommand {\CHII}     {C_{\rm HII}}
\newcommand {\fesc}     {f_{\rm esc}}
\newcommand {\taue}     {\tau_{\rm e}}
\newcommand {\nHbar}    {\bar{n}_{\rm H}}
\newcommand {\nHII}     {n_{\rm HII}}
\newcommand {\trec}     {\bar{t}_{\rm rec}}
\newcommand {\nidot}    {\dot{n}_{\rm ion}}
\newcommand {\zetai}    {\zeta_{\rm ion}}
\newcommand {\MUV}      {M_{\rm UV}}
\newcommand {\ML}       {$M$-$L$\ }

\newcommand {\Mvir}     {M_{\rm h}}
\newcommand {\pstar}    {\phi_\star}
\newcommand {\Lstar}    {L_\star}
\newcommand {\Mstar}    {M_{{\rm UV},\star}}
\newcommand {\Mpsi}     {M_{{\rm UV},\psi}}
\newcommand {\Lpsi}     {L_\psi}
\newcommand {\dndlnM}   {dn/{d\rm ln}\Mvir}
\newcommand {\Scut}     {S_{\rm cut}}
\newcommand {\Ocut}     {O_{\rm cut}}
\newcommand {\zs}       {z_{\rm start}}
\newcommand {\ze}       {z_{\rm end}}
\newcommand {\khf}      {k_{1/2}}
\newcommand {\kJeq}     {k_{J,{\rm eq}}}

\newcommand {\sref}[1]  {Section~\ref{#1}}
\newcommand {\fref}[1]  {Figure~\ref{#1}}

\newcommand {\eref}[1]  {Equation~(\ref{#1})}

\newcommand {\be}       {\begin{equation}}
\newcommand {\ee}       {\end{equation}}

\begin{document}

\title{Contrasting Galaxy Formation from Quantum Wave Dark Matter, $\psiDM$,
with $\Lambda$CDM, using Planck and Hubble Data}

\author{Hsi-Yu Schive\altaffilmark{1}, Tzihong Chiueh\altaffilmark{1,2,3},
Tom Broadhurst\altaffilmark{4,5}, \& Kuan-Wei Huang\altaffilmark{1}}

\altaffiltext{1}{Department of Physics, National Taiwan University, 10617 Taipei, Taiwan}
\altaffiltext{2}{Institute of Astrophysics, National Taiwan University, 10617 Taipei, Taiwan}
\altaffiltext{3}{Center for Theoretical Sciences, National Taiwan University, 10617 Taipei, Taiwan}
\altaffiltext{4}{Department of Theoretical Physics, University of the Basque Country UPV/EHU, E-48080 Bilbao, Spain}
\altaffiltext{5}{Ikerbasque, Basque Foundation for Science, E-48011 Bilbao, Spain}

\email{hyschive@ntu.edu.tw}
\shorttitle{High-z Galaxies in $\psiDM$}
\shortauthors{H-Y. Schive et al.}

\begin{abstract}
The newly established luminosity functions of high-z galaxies at
$4 \lesssim z \lesssim 10$ can provide a stringent check on dark matter
models that aim to explain the core properties of dwarf galaxies.
The cores of dwarf spheroidal galaxies are understood to be too large to be
accounted for by free streaming of warm dark matter without overly suppressing
the formation of such galaxies. Here we demonstrate with cosmological simulations that wave dark matter, $\psiDM$,
appropriate for light bosons such as axions, does not suffer this
problem, given a boson mass of $m_{\psi} \ge 1.2 \times 10^{-22} \eV$
($2\sigma$). In this case, the halo mass function is suppressed below
$\sim 10^{10} \Msun$ at a level that is consistent with the high-z
luminosity functions, while simultaneously generating the kpc-scale cores
in dwarf galaxies arising from the solitonic ground state in $\psiDM$.
We demonstrate that the reionization history in this scenario is consistent with the Thomson
optical depth recently reported by Planck, assuming a reasonable ionizing photon production rate.
We predict that the luminosity function should turn over slowly around
an intrinsic UV luminosity of $\MUV \gtrsim -16$ at $z \gtrsim 4$.
We also show that for galaxies magnified $\mathord{>}10\times$ in the Hubble Frontier Fields,
$\psiDM$ predicts an order of magnitude fewer detections than cold dark
matter at $z \gtrsim 10$ down to $\MUV \sim -15$,
allowing us to distinguish between these very different interpretations for
the observed coldness of dark matter.
\end{abstract}


\section{INTRODUCTION}
\label{sec:Intro}
 
The transitional stage when density perturbations first collapse to become
galaxies is now being reached with data of unprecedented depth. New
polarization measurements with the Planck satellite indicate an optical
depth of only $\taue=0.066\pm0.016$, for electron scattering of the
cosmic microwave background (CMB), which translates into an approximate
instantaneous reionization redshift of
$z \sim 8.8^{+1.7}_{-1.4}$ \citep{Planck2015}.

This may be compared with the ionization state of hydrogen absorption lines
revealed by distant, bright quasi-stellar objects (QSOs) and gamma-ray bursts
(GRBs), which has demonstrated that inter-galactic hydrogen has remained
highly ionized since at least $z \sim 6$. Beyond which evidence of a patchy
opacity is claimed in some QSO and GRB spectra
\citep{Chornock2014,Melandri2015,Becker2015} with a small mean enhancement
in the average neutral fraction by $z=6.5$ \citep{Hartoog2014}.
The forest at $z>6.5$ is not yet measured with any bright source, but a large
jump in neutral hydrogen fraction may be implied in the range $6.5<z<7.0$
from the statistical absence of strong Lyman-$\alpha$ emission from the most
distant galaxies amenable to spectroscopy \citep{Pentericci2014,Stark2015}.
Exceptions have been found with strong Lyman-$\alpha$ emission
at $z \sim 7.7$ \citep{Oesch2015},
and also the currently highest redshift galaxy established by spectroscopy
at $z \sim 8.7$ \citep{Zitrin2015}.
A relatively sharp reionization transition is
considered likely if galaxies are the dominant source of reionization, based
on advanced hybrid techniques \citep{Mesinger2015} which justifies the
instantaneous reionization redshift approximation \citep{Dijkstra2014,Mitra2015}.
More detailed 3D modeling of this transition remains very challenging when incorporating all relevant
processes which may affect ultraviolet (UV) ionization
\citep[e.g.,][]{Springel2003,Bromm2011,Wise2014}, including early galactic outflows
\citep[e.g.,][]{Frye2002} and their feedback effects
\citep[e.g.,][]{Scannapieco2001,Pieri2007,Booth2012}.

This relatively low value of $\taue$ implies that galaxies may not be expected
to be found in abundance at redshifts much higher than $z \sim 9$. Indeed
beyond this redshift only a handful of galaxies are claimed in the deepest
fields, among which the current most distant galaxy has $z \sim 10.7$,
which is discovered in the CLASH program and is highly magnified by
gravitational lensing \citep{Coe2013}.
Two other reliably estimated high-z galaxies are also known behind the new,
highly magnifying Hubble Frontier Field (HFF) clusters at $z \sim 9.6$
\citep{Zheng2012} and $z \sim 9.8$ \citep{Zitrin2014}. Nothing beyond this
has been found in HFF so far, despite the high magnifications by these
clusters \citep{Zitrin2009,Lam2014,Diego2015} and the great depth of these images in the near infrared, with
their potential to detect even higher redshift galaxies to a photometric
limit of $z \sim 11.5$ \citep{Coe2015}.

The UV luminosity function (LF) of distant galaxies is now well constructed
at $z \sim 4 - 10$ relying mainly on dropout galaxies detached in deep
field searches \citep[e.g.,][hereafter B15b]{Bouwens2015b}. The LF is seen to steadily
evolve to low number densities of high redshift galaxies and to steepen at
the faint-end slope as it does so (B15b). Evolution is also seen in terms of the
mean sizes of dropout galaxies which steadily decrease with increasing
redshift \citep{Bouwens2003,Holwerda2015}.
Currently, the behavior of the UV luminosity density at $z>8$ is hotly debated
with evidence of an accelerated decline at $z>8$ by B15b but with a counter claim by
\citet{McLeod2015} for the HFF with a reliance on parametric lens modeling.
This latter claim is at odds with \citet{Coe2015} who provide some
evidence of a deficit at high redshifts for the HFF, based on the first two
completed clusters. Of course in this $z \gtrsim 9$ redshift range data is
restricted to fewer detections in only infrared passbands lying close to
magnitude limits, so that crucial conclusions regarding galaxy formation are
still uncertain.

In this paper, we examine the high-redshift galaxy formation in the context of
a wave dark matter model, known as \emph{$\psiDM$}
\citep[][hereafter SCB14a]{Schive2014a} or \emph{fuzzy} dark matter
\citep{Hu2000}. In this scenario, dark matter is
assumed to be composed of extremely light bosons, such as axion-like
particles proposed by string theory \citep{Arvanitaki2010} or non-QCD axions
\citep{Chiueh2014}. They are non-thermally generated and can be described by
a single coherent wave function \citep{Turner1983,Goodman2000,Bohmer2007,Sikivie2009,Davidson2015,Guth2015}.
When self-interaction is negligible, the evolution of $\psiDM$ is governed
by the Schr\"{o}dinger-Poisson equation
\citep{Ruffini1969,Seidel1990,Widrow1993,Hu2000,Woo2009,Schive2014b}, with a single free
parameter, $m_{\psi}$, the dark matter particle mass.

The most prominent feature in $\psiDM$ is that the uncertainty principle
counters gravity below a Jeans scale, resulting in a
suppression of halos below $\sim 10^{10} \Msun$ and a flat density profile
within $\sim 0.1-1.0 \kpc$ of the centers of galaxies, assuming
$\PM \sim 1.0$ where $\PM = m_{\psi}/10^{-22}\eV$
\citep{Khlopov1985,Peebles2000,Hu2000,Matos2001,Lee2010,Marsh2010,Schive2014a}.
This boson mass scale can naturally arise in a non-QCD axion model
\citep{Chiueh2014}, lending support for the very light boson.
The $\psiDM$ model has become a viable dark matter candidate 
\citep[e.g.,][]{Woo2009,Mielke2009,Chavanis2011,Suarez2011,Robles2012,
Marsh2014,Rindler2014,Lora2014,Bray2014,Suarez2014,Bozek2015,Marsh2015a,
Martinez2015,Guzman2015,Madarassy2015,Harko2015},
especially given the increasingly strict limits of non-detections of the
weakly interacting massive particles (WIMPs) in the standard cold dark matter
\citep[CDM,][]{Akerib2014}. Various observable properties of this model
have been proposed
\citep{Amendola2006,Arvanitaki2011,Schive2014a,Schive2014b,Khmelnitsky2014,
Hlozek2015,VanTilburg2015,Stadnik2015}.

The first high-resolution cosmological simulations for the $\psiDM$ model
have recently generated exciting results (SCB14a).
We have directly demonstrated that indeed the large-scale structure of
$\psiDM$ is statistically indistinguishable from CDM, but differs radically
on small scales, where $\psiDM$ halos
form central solitonic cores surrounded by fine-scale, large-amplitude
granular textures. By applying a Jeans analysis to the stellar phase-space
distribution in the Fornax dwarf spheroidal (dSph) galaxy, which is known
to have a distinct core \citep[e.g.,][]{Amorisco2013}, we determine
$\PM=0.8\pm0.2$ ($1\sigma$), thereby providing the crucial normalization of this model
(SCB14a).

From our numerical simulations and theoretical arguments based on the
scaling symmetry of the Schr\"{o}dinger-Poisson equation and the uncertainty principle,
we subsequently derived a unique core-halo mass relation in $\psiDM$ \citep[][hereafter S14b]{Schive2014b},
$M_c \propto (1+z)^{1/2}\Mvir^{1/3}$, where $M_c$ and $\Mvir$ are the core
mass and halo mass, respectively, and $z$ is redshift.
This relation predicts that massive galaxies with $\Mvir \sim 10^{12} \Msun$ at
$z \sim 8$ will have compact solitons of $M_c \sim 10^9 \Msun$ within
$\sim 60 \pc$. Our simulations show that these dense solitonic cores form
promptly after halo collapse, and thus may help to explain the early onset
of QSO activity \citep{Trakhtenbrot2015} by acting as a massive focus for
gas accretion.

Another key prediction of $\psiDM$ is that galaxy formation is \emph{delayed}
relative to CDM because of the inherent Jeans scale. The preliminary results
of SCB14a showed that the first galaxies form at $z \sim 13$ with
$\Mvir \sim 10^9 - 10^{10} \Msun$, assuming $\PM \sim 1.0$ fixed by the scale
of dSph galaxy cores as described above. Halos below $\sim 10^9 \Msun$ are
significantly suppressed. We stress that the particle mass, $\PM$, is the only
free parameter here
assuming that the dark matter is made entirely of $\psiDM$ (see
\citealt{Marsh2014} for a mixed CDM and $\psiDM$ model).
The smaller the $\PM$, the greater the
difference between CDM and $\psiDM$.
Here we aim to establish whether a similar particle mass ($\PM \sim 1.0$) can
satisfy both the observed properties of dSph galaxies and the constraints
from high-z observations, such as galaxy counts, reionization history, and
Lyman-$\alpha$ forest, although some tension seems to exist \citep{Bozek2015}. It
is a well known issue for warm dark matter (WDM), usually termed as the
\emph{Catch 22} problem \citep{Maccio2012}, where the kpc-scale cores in
dSph galaxies require too small a WDM particle mass that
is in contradiction with high-z observations
\citep{Schneider2014,Schultz2014,Lovell2014}. Since the relation between core radius
and power spectrum suppression are different in $\psiDM$ and WDM
\citep{Hu2000,Schive2014b,Marsh2015a}, in this work we examine in detail
whether $\psiDM$ is clear of this serious problem facing WDM.

In this paper we conduct cosmological simulations to study the evolution of
halo mass function (MF) in the $\psiDM$ scenario, and connect it to the
recently established galaxy UV LF at $4 \lesssim z \lesssim 10$.
We explore the results of different $\psiDM$ particle masses ranging
from $\PM = 0.8$ to $3.2$. We predict the evolution of LF beyond the current
observational limit as a future test to distinguish between CDM and
$\psiDM$. We also perform analytic calculations to study the reionization
history in this context, and compare it to the Thomson optical depth recently
reported by Planck.
All magnitudes in this paper are quoted in the AB system
\citep[$M_{\rm AB}$,][]{Oke1983}.

The paper is structured as follows. In \sref{sec:Simulations} we describe
our simulation setup, including initial conditions and simulation
characteristics. We show the $\psiDM$ halo mass function in \sref{sec:MF},
and compare it with observations in \sref{sec:Predictions}. Finally, we
discuss and summarize our results in \sref{sec:Discussion}.

\section{SIMULATIONS}
\label{sec:Simulations}

In this section, we describe the initial power spectra and other
characteristics of our simulations for the study of the evolution of the
$\psiDM$ halo MF at high redshifts.

\subsection{Initial Power Spectra}
\label{subsec:InitPS}

The suppression of $\psiDM$ linear density power spectrum relative
to CDM can be expressed as
\be
P_{\psiDM}(k,z) = T_{\psiDM}^2(k,z)P_{\rm CDM}(k,z),
\label{eq:TranFunc}
\ee
where $P$ denotes the power spectrum and $T_{\psiDM}$ is the $\psiDM$ transfer
function
(strictly speaking it is the ratio between the transfer functions of $\psiDM$
and CDM).
In general $T_{\psiDM}$ is both redshift- and scale-dependent since
the balance between gravity and quantum pressure introduces a redshift-dependent
Jeans scale, $k_J(z)$, below which the structures cannot grow.
However, for the particle masses, redshift range, and halo masses of interest
in this work ($\PM \sim 1$, $z\sim 4-10$, $\Mvir \gtrsim 1\times10^9 \Msun$),
$T_{\psiDM}$ can be approximated as redshift-independent, as we demonstrate
below.

The redshift evolution of $\psiDM$ density perturbations during the
matter-dominated epoch can be described analytically by \citep{Woo2009}
\be
\rho_k(k,z) = A(k)\frac{3\cos\theta - \theta^2\cos\theta + 3\theta\sin\theta}{\theta^2},
\label{eq:GrowingMode}
\ee
where $\rho_k$ is the spatial Fourier component of the comoving density
perturbations, A(k) is the normalization constant,
$\theta(k,z) = \hbar k^2 \sqrt{1+z}/m_{\psi}H_0\sqrt{\Omega_{m0}}$,
$H_0$ is the present Hubble parameter, and $\Omega_{m0}$ is the present
matter density parameter. Setting $\theta^2=6$ gives the Jeans scale,
\be
k_J(z) \approx 69.1\,\PM^{1/2} \left( \frac{\Omega_{m0}h^2}{0.14} \right)^{1/4} (1+z)^{-1/4} \Mpc^{-1},
\label{eq:JeansK}
\ee
where $h$ is the dimensionless Hubble constant. For $k \ll k_J$, we have
$\rho_k \propto (1+z)^{-1}$, and thus $\psiDM$ grows like CDM; while for
$k \gg k_J$ the perturbations oscillate as $\rho_k \propto \cos\theta$.
Note that $k_J(z)$ increases slowly with time, and hence an oscillating mode
may become a growing mode at lower redshifts, but not vice versa.

\begin{figure}[t!]
\centering
\includegraphics[width=8.5cm]{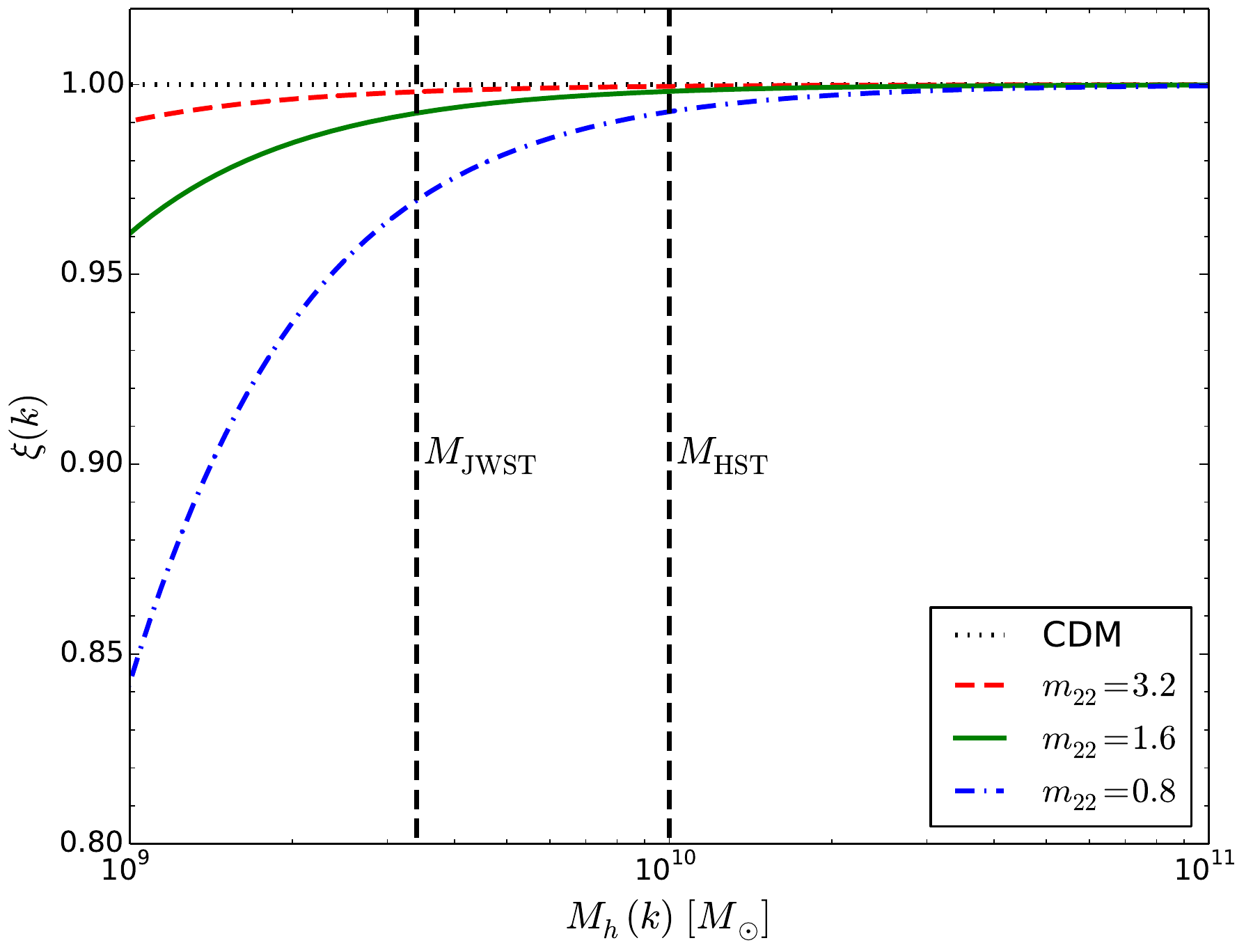}
\caption{
Growth rate ratio between $\psiDM$ and CDM density perturbations
(Eq. [\ref{eq:EvolvePS}]) during $30 \ge z \ge 4$. 
Vertical dashed lines highlight the mass of the faintest galaxies currently observed by HST at
$z\gtrsim 4$ ($M_{\rm HST}$) and that expected for JWST assuming a limiting
absolute magnitude of $\MUV=-15$ at $z=6$ ($M_{\rm JWST}$). Here we
have adopted the mass-luminosity relation described in \sref{subsubsec:CLF}.
Note that even for $\PM=0.8$ the growth rate ratios still reach
$\xi \sim 0.99$ at $\Mvir = M_{\rm HST}$ and $\xi \sim 0.97$ at
$\Mvir = M_{\rm JWST}$, indicating that the additional suppression of $\psiDM$
halos above $M_{\rm JWST}$ during this redshift interval of interest is almost
negligible (see text for details).
}
\label{fig:EvolvePS}
\end{figure}

To quantify the difference in growth rate between $\psiDM$ and CDM density
perturbations during a given redshift interval, $\zs \ge z \ge \ze$, we define
\be
\xi(k) = \frac{\rho_k(k,\ze)/\rho_k(k,\zs)}{\rho_k(k_0,\ze)/\rho_k(k_0,\zs)},
\label{eq:EvolvePS}
\ee
with $k_0 \ll k_J$ so that the denominator represents the growth in CDM.
For $k \ll k_J$, $\psiDM$ grows like CDM, and thus $\xi \sim 1$.
For $k \sim k_J$, quantum pressure starts to counter gravity and leads to
$\xi < 1$, indicative of `additional' suppression of $\psiDM$ halos during
this epoch.

\fref{fig:EvolvePS} shows $\xi(k)$ for various $\psiDM$ particle
masses. We take $\zs=30$ so that the $\psiDM$ density perturbations are still
in the linear regime, and $\ze=4$ to bracket the redshifts of interest in
this work. We convert the wavenumber to the halo virial mass via
$\Mvir = 4\pi(\pi/k)^3\rho_m/3$, where $\rho_m$ is the comoving matter density.
A relatively large deviation from CDM is found at the low-mass end for
a smaller particle mass because of the corresponding longer Jeans wavelength.
However, note that the faintest galaxies currently observed by the
Hubble Space Telescope (HST) at $z \gtrsim 4$ have $\Mvir \sim 10^{10} \Msun$
(see Sec. \ref{subsubsec:CLF} for the mass-luminosity relation adopted), at
which $\xi \sim 0.99$ for $\PM=0.8$. Even for the James Webb Space Telescope
\citep[JWST,][]{Gardner2006} assuming a limiting absolute magnitude of
$\MUV \sim -15$ at $z \sim 6$, we still have $\xi \sim 0.97$ for $\PM=0.8$.
It demonstrates that \emph{for the particle masses, redshifts, and halo masses
of interest when comparing with current and forthcoming observations,
(i) the growth rate of linear density power spectra
in CDM and $\psiDM$ are similar, and (ii) the $\psiDM$ transfer function
$T_{\psiDM}$ can be well approximated as redshift-independent}.
This is primarily due to that the Jeans mass at $z=30$ for $\PM=0.8$ is
$\sim 2.7\times10^8 \Msun$, well below the observational limits.
Note, also, that the smallest halos resolved in our simulations have
$\Mvir \sim 3\times10^8 \Msun$, close to the Jeans mass and hence
$\xi \sim 0.52$ for $\PM=0.8$. Therefore the halo MF at the
low-mass end may be, in this sense, slightly underestimated in our simulations.

\begin{figure}[t!]
\centering
\includegraphics[width=8.5cm]{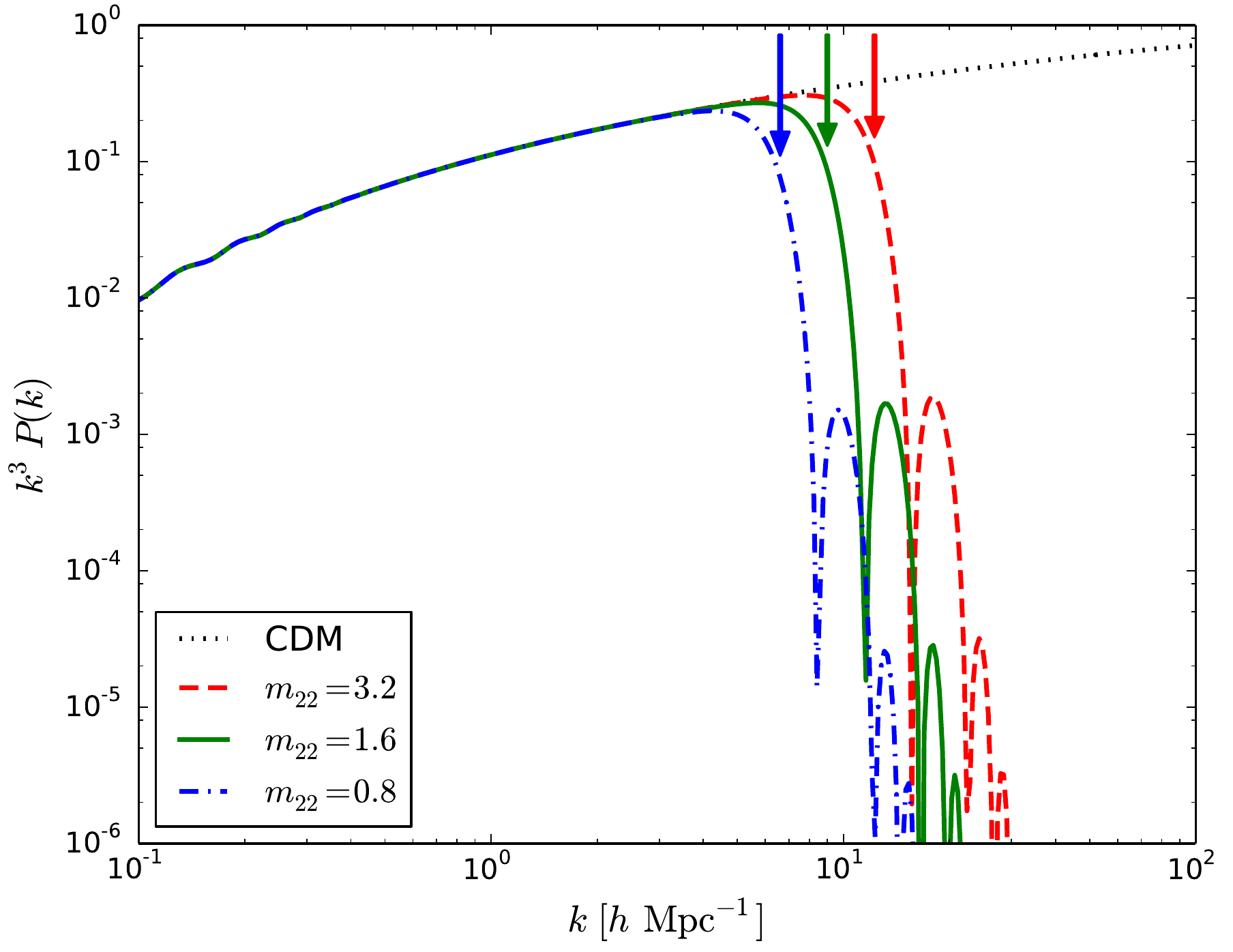}
\caption{
Linear matter power spectra at $z=30$ in CDM and $\psiDM$.
$\psiDM$ power spectra are obtained using Equations (\ref{eq:TranFunc})
and (\ref{eq:TranFuncHu}),
where we have assumed that $T_{\psiDM}$ can be well approximated as
redshift-independent during $30 \ge z \ge 0$ (see Fig. \ref{fig:EvolvePS}).
Note that the smaller the particle mass, the stronger
the suppression at the high-k end. Arrows indicate the half-mode wavenumbers,
$\khf$ (Eq. [\ref{eq:HalfMode}]), where the power spectra drop by a factor
of four compared to CDM.
}
\label{fig:InitPS}
\end{figure}

The $\psiDM$ transfer function at $z=0$ is given by \citep{Hu2000}
\be
T_{\psiDM} \approx \frac{\cos x^3}{1+x^8},\;\; x=1.61\,\PM^{1/18}\frac{k}{\kJeq},
\label{eq:TranFuncHu}
\ee
which is assumed to be redshift-independent relevant for the redshifts
and wavenumbers of interest. Here $\kJeq=9\,\PM^{1/2}\Mpc^{-1}$
is the Jeans wavenumber at matter-radiation equality.
\fref{fig:InitPS} presents the linear matter power spectra at $z=30$,
which exhibit sharp breaks at $k \sim \kJeq$
and strong oscillations for $k > \kJeq$.
We can also define the characteristic scale to be the `half-mode' scale,
$\khf$, where $T_{\psiDM}(\khf) = 1/2$. \eref{eq:TranFuncHu}
then gives
\be
\khf \approx 5.1\,\PM^{4/9}\Mpc^{-1},\;\;
M_{1/2} \approx 3.8\times10^{10}\,\PM^{-4/3} \Msun,
\label{eq:HalfMode}
\ee
where $M_{1/2} = 4\pi(\pi/\khf)^3\rho_m/3$ is the characteristic halo mass where
a noticeable difference between $\psiDM$ and CDM MFs is expected.
Note that the strong suppression at $k \sim \kJeq$ shown by \eref{eq:TranFuncHu}
is mainly determined during the radiation-dominated epoch \citep{Hu2000},
and thus cannot be solely explained by \eref{eq:GrowingMode} which is only
valid during the matter-dominated epoch.

\subsection{Simulation Characteristics}
\label{subsec:SimuChar}

Bona-fide $\psiDM$ simulations involve solving the Schr\"{o}dinger-Poisson
equation, which is extremely challenging owing to its wave nature. The matter
wave dispersion relation, $\omega \propto \lambda^{-2} \propto v^2$, where
$\omega$, $\lambda$, $v$ are the angular frequency, wavelength, and
velocity, respectively, indicates that exceptionally high spatial and
temporal resolutions are required for resolving the wave functions of
high-velocity flows throughout a simulation box (SCB14a).
Numerically, we find that a comoving spatial resolution as high as
$\sim 1 \kpch$ is required to properly resolve a flow with a moderate
peculiar velocity of $\sim 100 \kms$ at $z \sim 13$. Otherwise we find
the flow velocity can be underestimated, leading to lower
mass accretion rate and underestimation of MF.
In the extreme case, a $\psiDM$ simulation in a $30 \Mpch$
box with a uniform $1 \kpch$ spatial resolution will consume $\sim 400$
terabytes of memory, which is impractical in any modern supercomputer.
As a result, even the state-of-the-art $\psiDM$ simulations currently can
only fully resolve a comoving box as small as $1.4 \Mpch$ to $z=0$
(SCB14a).

In this paper we mainly focus on determining the $\psiDM$ halo MF
above $\sim 1\times10^9 \Msun$ at $z \ge 4$, for which most halos are
isolated and thus insensitive to the subtle differences
between CDM and $\psiDM$ halos. Nor are we interested here in the complex
wave nature of the internal density profiles of the halos, which we have
already established in our previous wave-based simulations (SCB14a, S14b).
As we demonstrated in the previous subsection,
the growth rates of density perturbations in CDM and $\psiDM$ are similar in
the context of this work and differ mainly in their initial amplitudes.
Moreover, S14b verified that CDM and $\psiDM$ halos have
similar virial masses during the same collapse process. All these facts
indicate that, \emph{for the purpose of this study, it is appropriate to use
simulations of collisionless particles with $\psiDM$ initial power
spectra to approximate real $\psiDM$ simulations}. This is the approach
adopted in this work, which is essentially the same as most WDM simulations
where initial thermal velocity are ignored. Real $\psiDM$ simulations
for supporting these arguments, solving either wave function directly or an
alternative fluid-like form \citep[e.g.,][]{Mocz2015,Marsh2015b}, are for future work.

All simulations are run from $z=100$ to 4. Since the linear power spectra relevant
for this study do not change in shape after $z=30$, we can directly apply
\eref{eq:TranFuncHu} to obtain the $\psiDM$ power spectra at $z=30$
(see Fig. \ref{fig:InitPS}). To capture the rare non-Gaussian
peaks, which are the seeds of first galaxies, due to nonlinearity set in
as early as $z \sim 30$, we then extrapolate the $z=30$ spectra to $z=100$ for
which the amplitude is $\sim 3.3$ times smaller to ensure all perturbations are
Gaussian. By doing so, the simulations of collisionless particles preserve
the shape of the spectra from $z=100$ to $z=30$ but allow for the
development of rare non-Gaussian peaks.

We perform simulations with the \textsc{GADGET-2} N-body code 
\citep{Springel2005}. We adopt the \textsc{CAMB} package \citep{Lewis2000} to
generate the CDM transfer function, and construct initial conditions for
simulations using the \textsc{MUSIC} code \citep{Hahn2011}. We adopt the
cosmological parameters consistent with the
WMAP9 data \citep{WMAP9}: $\Omega_{m0}=0.284$, $\Omega_{\Lambda0}=0.716$,
$h=0.696$, $\sigma_8=0.818$, and $n_s=0.962$. We choose three different
simulation configurations with $(L,N)=(15\Mpch,\,512^3)$,
$(15\Mpch,\,1024^3)$, and $(30\Mpch,\,1024^3)$, where $L$ is the comoving
box size and $N$ is the total number of simulation particles. The
corresponding simulation particle masses are $\sim 2.8\times10^6 \Msun$ and
$\sim 3.6\times10^5 \Msun$ for the lower and higher mass-resolution
simulations, respectively. For each simulation configuration, we run four
different dark matter models: CDM, $\psiDM$ with $\PM=0.8$, 1.6, and 3.2.

\section{MASS FUNCTION}
\label{sec:MF}

The main aim of our simulations is to determine the halo MF
as a function of $\psiDM$ particle mass. Intuitively, a sharp break in the
initial power spectrum should translate into a strong suppression of
low-mass halos, as verified by the Sheth-Tormen \citep[][hereafter ST99]{Sheth1999} MF
with a sharp k-space window function \citep{Schneider2013}. However, it is
well known that the particle simulations with an initial power spectrum
cut-off suffer from the formation of spurious halos, especially at
low masses \citep{Wang2007,Angulo2013,Schneider2013}.

\begin{figure}[t!]
\centering
\includegraphics[width=8.5cm]{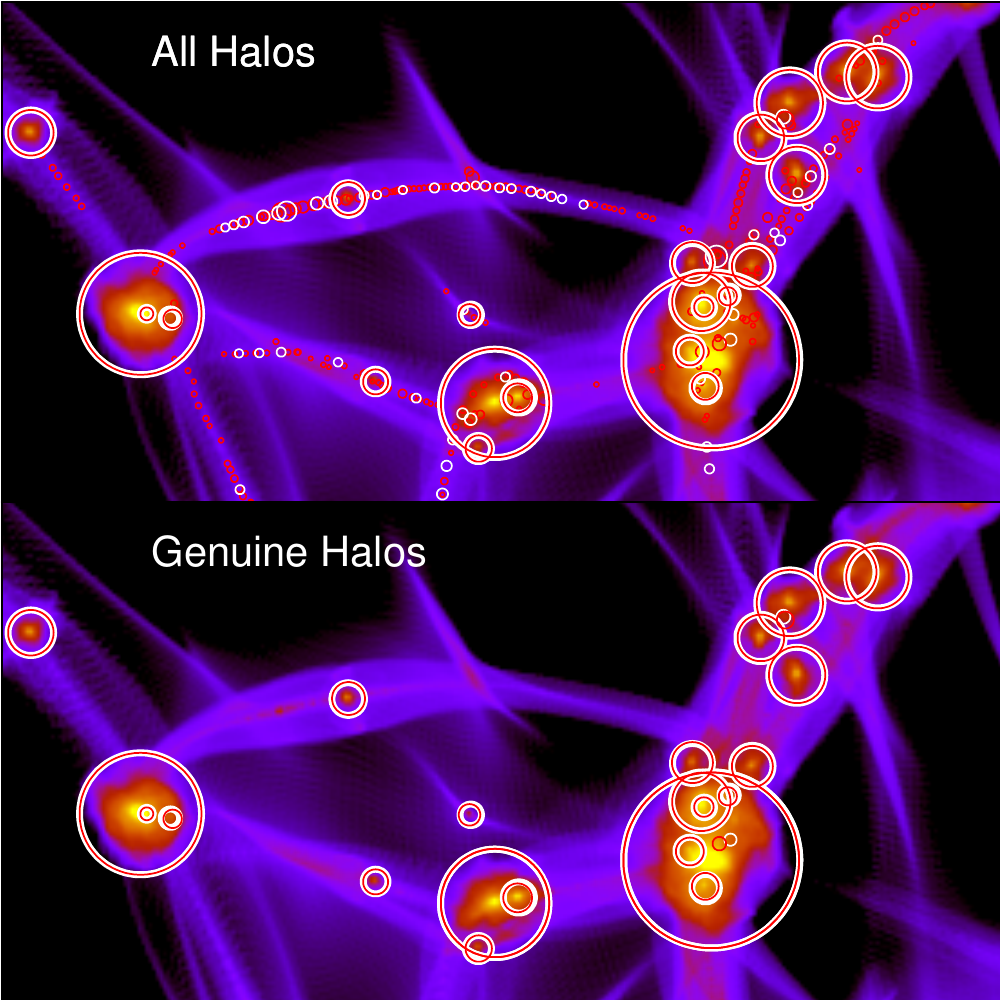}
\caption{
Density field at $z=4$ for $\psiDM$ simulations with $\PM=1.6$ in a
$15 \Mpch$ box. Each image displays a projected field for a $3 \Mpch$ thick
slab with a size of $2.70 \times 1.35 \Mpch$.
White and red circles show halos more massive than $2\times10^7 \Msun$
in the $512^3$ and $1024^3$ simulations, respectively, where the radii
of circles equal the halos' virial radii. The most massive halo has a mass
of $\sim 1\times10^{12} \Msun$. The upper panel shows both genuine and
spurious halos, while the lower panel only shows genuine halos. Suspicious
low-mass halos, which are mostly confined along filaments and have no clear
counterparts in the $512^3$ and $1024^3$ runs, are identified as spurious,
while only massive halos with a good match between low- and high-resolution
simulations are regarded as genuine.
}
\label{fig:HaloMap}
\end{figure}

\begin{figure}[t!]
\centering
\includegraphics[width=8.5cm]{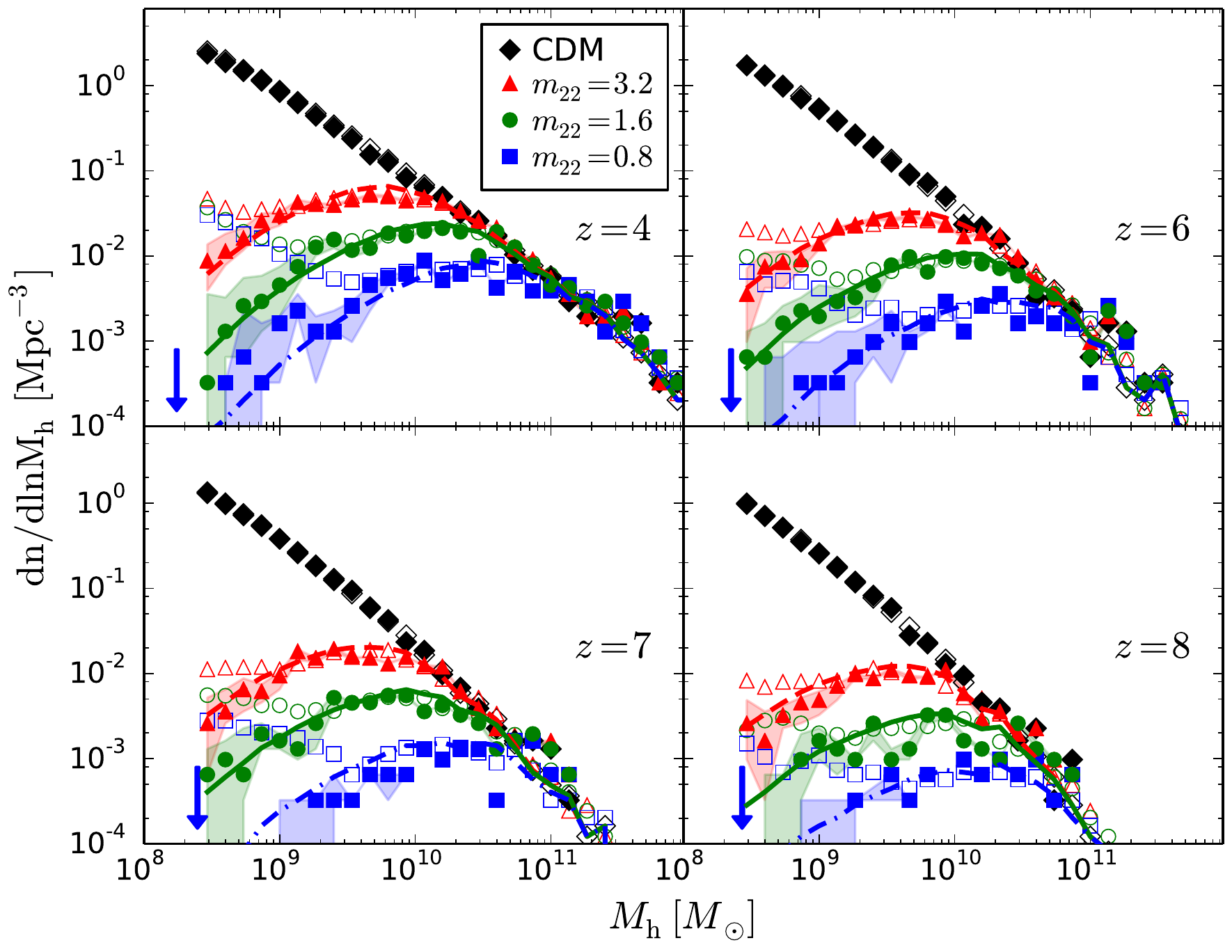}
\caption{
Halo mass function (MF) in logarithmic mass bins. The open symbols
represent the original MF containing both genuine and spurious
halos in the $30 \Mpch$ simulations with $1024^3$ particles, and the
filled symbols show the genuine MF with spurious halos removed in the
$15 \Mpch$ simulations using $512^3$ and $1024^3$ particles for
estimating the spatial overlap factor. The spurious halos outnumber genuine halos
at low masses, resulting in an unphysical upturn at the low-mass end of
the original MF, especially for lower $\PM$. By contrast, the genuine
MF reveals a prominent drop at the low-mass end, as anticipated and seen in our
high-resolution wave-based simulations, reported earlier. The shaded regions
indicate the uncertainties of genuine MF by varying $\Scut$ and $\Ocut$
by $\pm 20\%$ (see Appendix \ref{sec:SpuriousHalo}). Various lines show the
analytic form, \eref{eq:MFFit}, which fit the simulation results well.
Arrows mark the minimum $\psiDM$ halo masses proposed by S14b
for $\PM=0.8$.
}
\label{fig:MF}
\end{figure}

These spurious halos are caused by artificial fragmentation due to numerical
artifacts \citep{Wang2007}, and are mostly confined along cosmic filaments (see
Fig. \ref{fig:HaloMap}, upper panel). They outnumber genuine halos
below a characteristic mass, which linearly depends on the mean interparticle
separation \citep{Wang2007}, resulting in a prominent upturn in MF at the
low-mass end (see Fig. \ref{fig:MF}, open symbols). We define `protohalo' as
the initial particle positions of an identified halo. \citet{Lovell2014} showed
that the protohalos of genuine and spurious halos have distinct features.
Genuine protohalos are spheroidal and have a good match between low- and
high-resolution simulations, while spurious protohalos have disc-like shapes
and their masses and positions are sensitive to the simulation resolution,
and thus do not have clear counterparts in simulations with different
resolution.
To identify and remove these artificial halos,
we thus adopt a similar approach suggested by \citet{Lovell2014} based
on the shape of the protohalos and the spatial overlap between
low-resolution protohalos and their high-resolution counterparts.
See Appendix \ref{sec:SpuriousHalo} for a more detailed description
of the algorithms adopted.

\fref{fig:MF} shows the halo MF obtained in our simulations. For comparison,
we show both the `original' MF (containing both genuine and spurious
halos) in the $30 \Mpch$ simulations with $1024^3$ particles, and the `genuine'
MF (with spurious halos removed) in the $15 \Mpch$ simulations using $512^3$
and $1024^3$ particles for estimating the spatial overlap factor.
The original $\psiDM$ MF shows a prominent upturn at the
low-mass end due to the contamination from spurious halos, especially for
lower $\PM$. By contrast, the genuine $\psiDM$ MF features a clear drop
at low masses for all redshifts and particle masses, apparently different
from CDM and in agreement with the expectation from a sharp break in the
$\psiDM$ initial power spectra. It is also consistent with the minimum
$\psiDM$ halo mass at $z \gtrsim 1$ proposed by S14b,
$M_{min} = 3.7\times10^7\,\PM^{-3/2}\,(1+z)^{3/4} \Msun$
(indicated by arrows in Fig. \ref{fig:MF} for $\PM=0.8$).
On the other hand, the original and genuine
CDM MFs are almost indistinguishable, which is no surprise since we assume
most CDM halos are genuine when calibrating the thresholds for removing
spurious halos.

The shaded regions in \fref{fig:MF} indicate the uncertainties of genuine
$\psiDM$ MF by varying $\Scut$ and $\Ocut$ by $\pm 20\%$ (see Appendix
\ref{sec:SpuriousHalo}).
It shows that the finding of strong suppression of low-mass halos in $\psiDM$
is reliable, but the exact slope at the low-mass end is still uncertain.
In the high-mass end ($\Mvir \gtrsim 10^{11} \Msun$)
the original MF is smoother because of a larger simulation box. Note, however,
that in the intermediate mass range
($\Mvir \sim 3\times10^9 - 1\times10^{11} \Msun$)
the original and genuine MFs are reasonably consistent with
each other, suggesting that in this mass range (i) most halos are genuine
and (ii) a $15 \Mpch$ simulation box is sufficient to obtain an accurate MF.
These results make the $\psiDM$ MF obtained in this work more robust
for the purpose of comparing with observations.

As shown in \fref{fig:MF}, the genuine $\psiDM$ MF can be well fitted by
the following analytic form:
\be
\left.\frac{dn}{d\Mvir}\right\rvert_{\psiDM}(\Mvir,z) =
\left.\frac{dn}{d\Mvir}\right\rvert_{{\rm CDM}}(\Mvir,z)
\left[ 1 + \left( \frac{\Mvir}{M_0} \right)^{-1.1} \right]^{-2.2},
\label{eq:MFFit}
\ee
where $dn/d\Mvir$ is the halo MF and $M_0=1.6\times10^{10}\,\PM^{-4/3} \Msun$
is the characteristic mass below which MF starts to drop noticeably.
CDM corresponds to $\PM \to \infty$.
The facts that $M_0$ has the same particle mass dependence as the
half-mode mass $M_{1/2}$ (Eq. [\ref{eq:HalfMode}]) and $\psiDM$ MF
drops by a factor of two relative to CDM at $\Mvir \sim M_{1/2}$
reinforce our simulation results.
Also note that the suppression term,
$( 1 + (\Mvir/M_0)^{-1.1})^{-2.2}$, is redshift-independent. It is
expected since in this work the effect of quantum pressure is taken into
account only for the initial conditions. In detail for $\psiDM$ the
suppression of low-mass halos will be redshift-dependent, but
the characteristic mass $M_0$ is still expected to be almost
redshift-independent since it is mainly determined during the
radiation-dominated epoch \citep{Hu2000}. We emphasize that the faintest
galaxies currently observed at $z \gtrsim 4$
have $\Mvir \sim 10^{10} \Msun$ (see Fig. \ref{fig:CumuMF} and
Sec. \ref{subsubsec:CLF}), which is close to $M_0$ and hence is insensitive to
the uncertainties at low masses of MF ($\Mvir \lesssim 10^9 \Msun$) caused by
neglecting the dynamical effect of quantum pressure
and the removal of spurious halos. \eref{eq:MFFit} thus provides a very
convenient comparison between models and observations (see next section).

\section{PREDICTIONS VS OBSERVATIONS}
\label{sec:Predictions}

The rest-frame UV LF at high redshifts have become increasingly well defined
and hence useful for testing a range of dark matter models in detail. The
physical mechanisms assumed to solve the small-scale issues of CDM in the
Local Group will likely suppress the formation of faint galaxies at high
redshifts as well,
so that it is important to examine whether too few high-z galaxies
are created and with a too small Thomson optical depth to CMB.
This is a well-known issue for WDM, usually termed as the \emph{Catch 22}
problem \citep{Maccio2012}. In this section we examine the level of
consistency in the case of $\psiDM$.

\subsection{Cumulative Mass Function}
\label{subsec:CumuMF}

The cumulative galaxy number density, defined as the total number of galaxies
per unit comoving volume at a given redshift, can be converted into a lower
limit of $\PM$, below which the $\psiDM$ MF cannot account for the observed
counts of galaxies. To relate the UV magnitude $\MUV$ of a 
galaxy to its corresponding halo mass, we first adopt the abundance
matching technique \citep{Vale2004} which equates the cumulative UV LF,
$\Psi(\mathord{<}\MUV,z)$, to the cumulative halo MF,
$n(\mathord{>}\Mvir,z)$. An alternative approach using the conditional LF
formalism will be discussed later.

For a given LF, one can apply abundance matching to either $\psiDM$ MF
with the particle masses of interest \citep{Bozek2015} or CDM MF
\citep{Schultz2014}, both of which have advantages and disadvantages. 
The former provides a model-independent constraint since it simply checks
whether the total numbers of $\psiDM$ halos at various redshifts are sufficient
to account for the observed counts, regardless of the underlying
mass-luminosity ($M$-$L$) relation. However, the inferred mass-to-light ratio
features a sharp, and probably unphysical, drop at the faint end
\citep{Bozek2015}. This approach therefore provides a more conservative
estimation of $\PM$. By contrast, the latter leads to a power-law \ML
relation at the faint end \citep{Schultz2014}, which is more plausible.
However, it fundamentally assumes that CDM matches the observed LF perfectly
and that $\psiDM$ follows exactly the same \ML relation as CDM, both of which may
not be necessarily true. Consequently, any suppression in the $\psiDM$ MF
translates directly into a deficit of galaxies, resulting in 
a higher, and likely overestimated, lower limit for $\PM$.

\begin{figure}[t!]
\centering
\includegraphics[width=8.5cm]{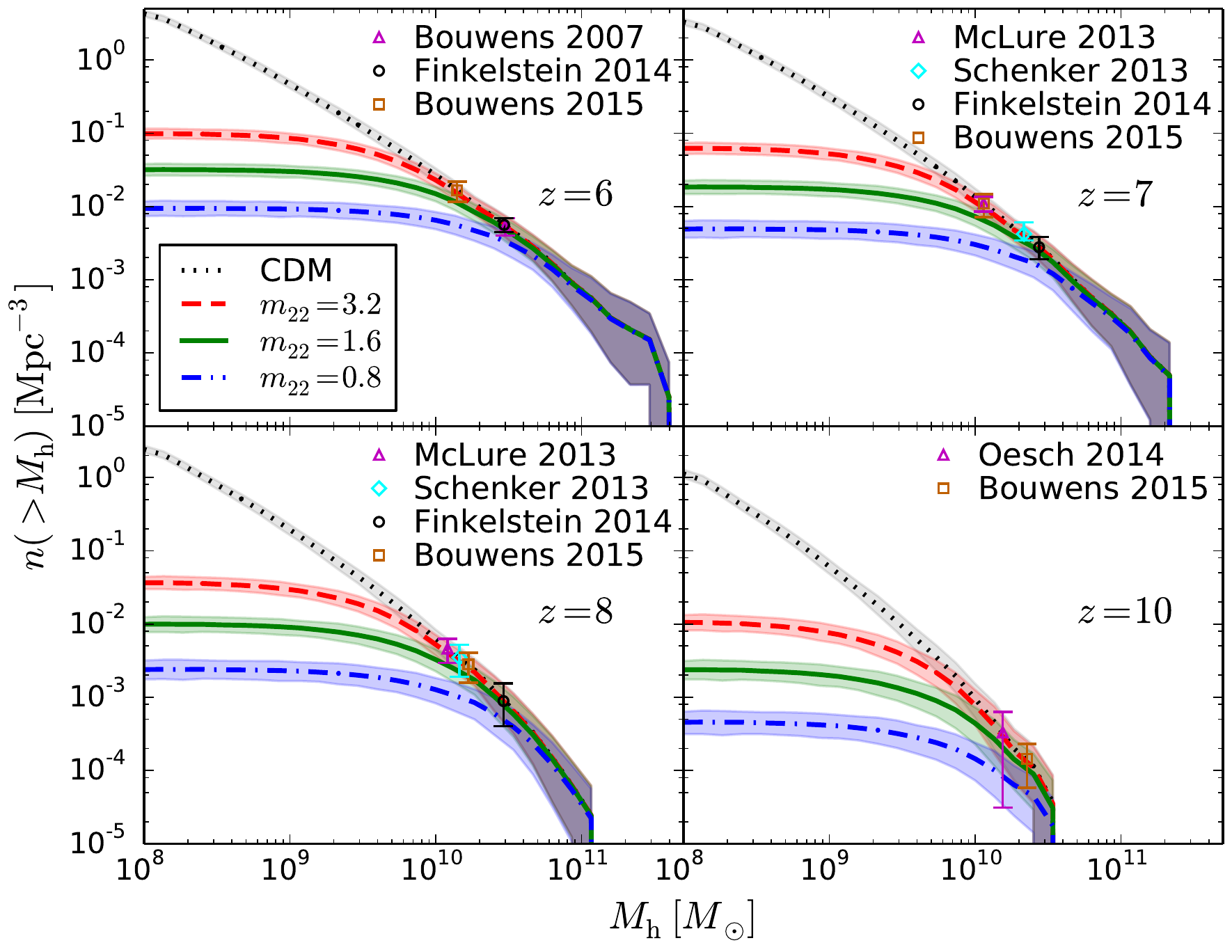}
\caption{Cumulative mass function (MF) at $z=6-10$. The shaded regions indicate
the $2\sigma$ uncertainties. Error bars show the
observational constraints ($2\sigma$ at $z=6-8$ and $1\sigma$ at $z=10$),
which match the CDM MFs perfectly simply because they are derived
from applying abundance matching to the CDM MFs.
Note that in this approach the faintest galaxies currently observed at
$z \sim 4-10$ all have $\Mvir \sim 10^{10} \Msun$.
$\psiDM$ cumulative MFs have finite upper limits due to the
strong suppression of low-mass halos. $\PM=3.2$ and $1.6$ are consistent
with the observations \citep{Bouwens2007,McLure2013,Schenker2013,Finkelstein2014,Oesch2014,Bouwens2015b},
while $\PM=0.8$ has insufficient halos with $\Mvir \sim 10^{10} \Msun$
at $z=6-8$ (see text for details).
}
\label{fig:CumuMF}
\end{figure}

\fref{fig:CumuMF} shows our cumulative MFs for both CDM and $\psiDM$.
CDM MF is constructed from a $30 \Mpch$ simulation
with $1024^3$ particles, and $\psiDM$ MF is obtained by
integrating \eref{eq:MFFit}. The $2\sigma$ uncertainties are estimated
using bootstrap resampling over 125 subvolumes, each with a side length of
$6 \Mpch$. Here the halo masses corresponding to various observational data
points are determined by applying abundance matching to the CDM MF, 
which essentially forces observations to be consistent with CDM.
In this approach, we find that the faintest galaxies currently observed at
$z \sim 4-10$ all have $\Mvir \sim 10^{10} \Msun$.
We demonstrate that $\psiDM$ with $\PM=3.2$ and $1.6$ are still consistent
with the observations, while $\PM=0.8$ does not have a sufficient number of
halos with $\Mvir \sim 10^{10} \Msun$ at $z=6-8$. Using the recent LF of
B15b leads to a lower limit of $\psiDM$ particle mass,
$\PM \ge 1.5$ ($2\sigma$). By contrast, 
to infer the results of applying abundance matching to the $\psiDM$ MF,
one can simply move the observational data points in \fref{fig:CumuMF}
toward smaller halo masses until, if possible, touching the $\psiDM$ cumulative
MF with a specific particle mass. This approach decreases the lower
limit to $\PM \ge 0.9$ ($2\sigma$), and hence significantly reduces the tension
between observational constraints and smaller $\psiDM$ particle masses.

We emphasize that the two estimations of $\psiDM$ particle mass given above,
namely, $\PM \ge 1.5$ and $\PM \ge 0.9$, are determined using two extreme
models of \ML relation, and therefore likely bracket the uncertainty
of the lower limit of $\PM$.
In fact, for $\psiDM$, it does not make much sense to apply the \ML relation
predicted by CDM from abundance matching. The lower limit $\PM \ge 1.5$ is therefore
likely overestimated.
In the next subsection we provide a more plausible
estimation of $\PM$ based on a less model-dependent \ML relation.

\subsection{Luminosity Function}
\label{subsec:LF}

\subsubsection{Conditional Luminosity Function}
\label{subsubsec:CLF}

As our preferred method to constrain the $\psiDM$ particle mass we adopt the
conditional LF model \citep{Cooray2005}, which has
been shown to be able to reproduce well the high-z UV LF in the context of CDM \citep{Bouwens2008,Bouwens2015b}.
The conditional LF, denoted as $\phi_c(L|\Mvir,z)$, describes the probability density of halos with
mass $\Mvir$ hosting galaxies with UV luminosity of $L$. It is modeled by
a lognormal distribution,
\be
\phi_c(L|\Mvir,z) = \frac{1}{\sqrt{2\pi}\ln(10)\Sigma L}
                     \exp\left\{ -\frac{\log[L/L_c(\Mvir,z)]^2}{2\Sigma^2} \right\},
\label{eq:CondLF_Main}
\ee
which has a dispersion of $\ln(10)\Sigma$ and peaks at $L_c(\Mvir,z)$,
the \ML relation of the central galaxy. Following B15b,
we parameterize $L_c$ as
\be
L_c(\Mvir,z) = L_0 \frac{(\Mvir/M_1)^{p}}{1+(\Mvir/M_1)^{q}}
               \left( \frac{1+z}{4.8} \right)^r,
\label{eq:CondLF_Lc}
\ee
where $M_1$ gives the characteristic halo mass. The \ML relation asymptotes
to $L_c \propto \Mvir^{p}$ when $\Mvir \ll M_1$ and $L_c \propto \Mvir^{p-q}$
when $\Mvir \gg M_1$. For a given $\phi_c$, the LF can then be calculated by
\be
\phi(L,z) = \int_{0}^{\infty} \phi_c(L|\Mvir,z) \frac{dn}{d\Mvir}(\Mvir,z)d\Mvir,
\label{eq:CondLF_Lc2L}
\ee
where $dn/d\Mvir$ is the halo MF.

\begin{table}
\setlength{\tabcolsep}{3pt}
\label{tab:CondLF}
\begin{center}
\caption{Parameters of the Conditional LF Model}
\begin{tabular}{lccccccc}
\hline
Model & $L_0$ & $M_1$ & $\Sigma$ & $p$ & $q$ & $r$ & $\chi_{red}^2$ \\
& ($M_{\rm AB}$) & ($\Msun$) \\
\hline
CDM                   & -20.7 & $2.7\times10^{11}$ & 0.16 & 1.6 & 1.2 & 1.9 & 1.4 \\
$\PM=3.2$             & -20.9 & $3.1\times10^{11}$ & 0.16 & 1.5 & 1.2 & 1.9 & 1.5 \\
$\PM=1.6$             & -21.1 & $4.0\times10^{11}$ & 0.16 & 1.4 & 1.2 & 1.9 & 1.9 \\
$\PM=0.8$             & -21.7 & $7.8\times10^{11}$ & 0.16 & 1.2 & 1.1 & 1.8 & 3.1 \\
B15b\tablenotemark{a} & -21.9 & $1.2\times10^{12}$ & 0.16 & 1.2 & 1.0 & 1.5 &     \\
\hline
\end{tabular}
\end{center}
\tablenotemark{a} \citet{Bouwens2015b}.
\end{table}

Given the above conditional LF formalism, we then use chi-square fitting
on the observed LF of B15b at $z=5-10$ to determine the parameter set
($L_0$, $M_1$, $\Sigma$, $p$, $q$, $r$). Table 1 shows the best-fit parameters
and the corresponding reduced chi-square ($\chi_{red}^2$). We fix $\Sigma$ to
$0.16$ both because it is not well constrained owing to the substantial uncertainties
at the bright end and because it does not influence the faint-end slope, which is
most important for constraining $\PM$.
Note that the faint-end \ML relation, $L_c \propto \Mvir^{p}$, is flatter for
smaller $\PM$
so as to compensate for the stronger suppression
of faint galaxies. In addition, note that from
\eref{eq:CondLF_Lc} the faintest galaxies currently observed at
$z \sim 4-10$ all have $\Mvir \sim 10^{10} \Msun$, consistent with the
results obtained by applying abundance matching to CDM
(see Fig. \ref{fig:CumuMF}).
A limiting absolute magnitude of $\MUV \sim -15$ at $z \sim 6$,
appropriate for JWST, corresponds to $\Mvir \sim 4\times10^9 \Msun$.

\begin{figure}[t!]
\centering
\includegraphics[width=8.5cm]{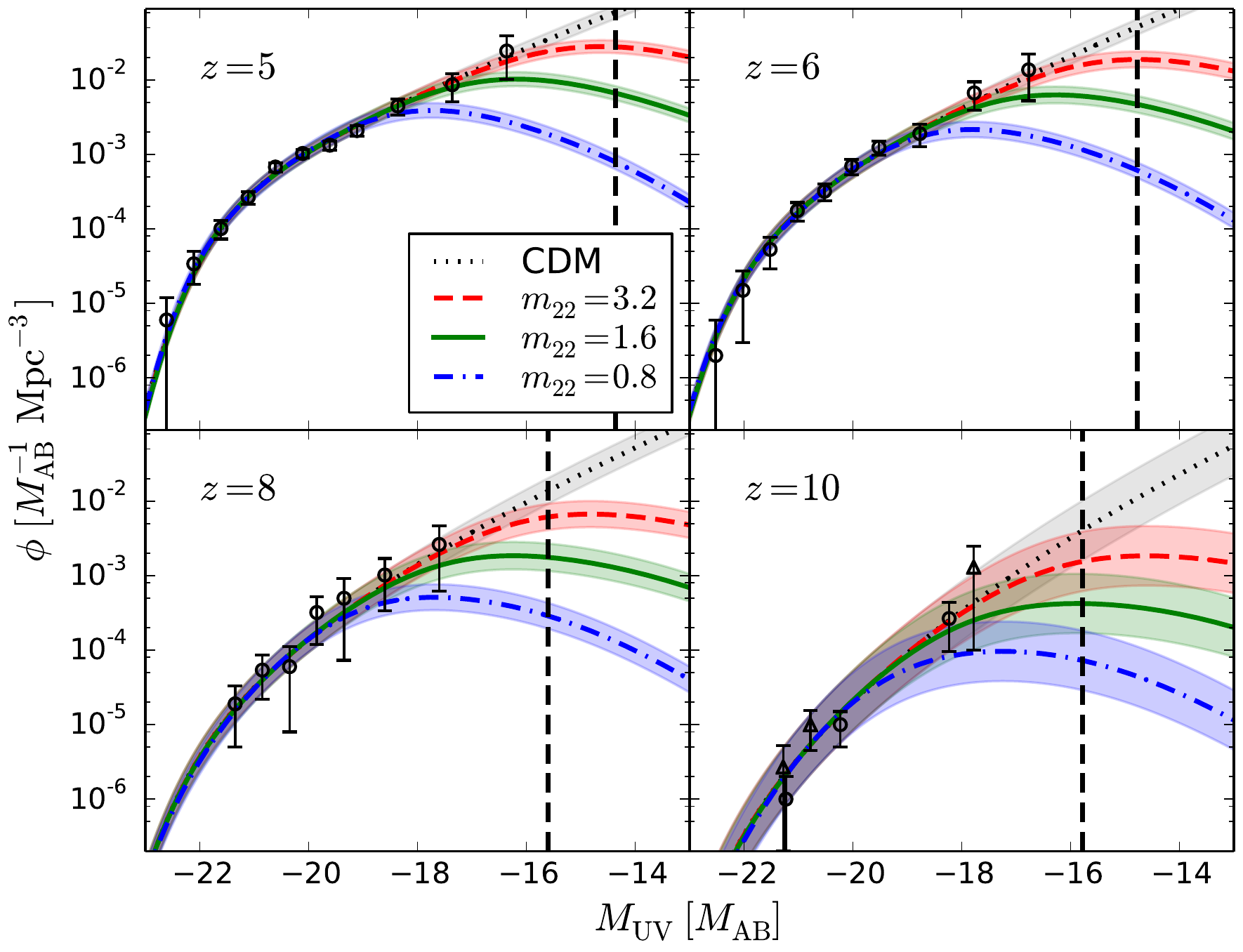}
\caption{Luminosity function (LF) at $z=5-10$ predicted by the
conditional LF model. The shaded regions indicate the $2\sigma$
uncertainties. Error bars show the observed LFs ($2\sigma$ at $z=5-8$ and $1\sigma$ at $z=10$)
of B15b (open circles) and \citet[][open triangles]{Oesch2014}.
$\psiDM$ LF shows a drop at the faint end due to the suppression of low-mass halos.
$\PM=3.2$ and $1.6$ are consistent with the observations,
while $\PM=0.8$ shows a deficit of faint galaxies with $\MUV \gtrsim -18$,
especially at $z \le 6$ (see text for details).
Vertical dashed lines highlight the limiting absolute magnitudes of JWST,
which are assumed to be two magnitudes fainter than those of HST.
}
\label{fig:CondLF}
\end{figure}

\fref{fig:CondLF} shows our predicted galaxy UV LF at $z=5-10$ using the conditional
LF formalism described above. We use the ST99 MF
for CDM, and combine it with the ratio given in \eref{eq:MFFit}
to get the $\psiDM$ MFs with various $\PM$.
We add $2\sigma$ variations estimated from the MF at $\Mvir \sim 10^{10} \Msun$
in our $30 \Mpch$ simulations to capture the uncertainties
of the predicted LF around the faintest LF bins of B15b.
The $\psiDM$ LF shows a clear decline at the faint end, which is distinctly
different from the CDM prediction and will be directly testable
with forthcoming observations such as JWST.
Note that this feature results from the assumption of a power-law \ML
relation at the faint end (Eq. [\ref{eq:CondLF_Lc}]), and thus cannot be
captured by the usual abundance matching.
The cases $\PM=3.2$ and $1.6$ are found to be consistent with the current observations,
while $\PM=0.8$ shows an apparent deficit of faint
galaxies with $\MUV \gtrsim -18$, especially at $z \le 6$.
This result is consistent with the analysis of cumulative MF
(see \fref{fig:CumuMF}). Using the faintest LF bins of
B15b (open circles in Fig. \ref{fig:CondLF}) leads to $\PM \ge 1.2$ ($2\sigma$).
Note that this constraint lies in-between
those obtained in \sref{subsec:CumuMF} ($\PM \ge 0.9$ and $1.5$), as expected,
since the faint-end \ML relation adopted here can be regarded as a
compromise between the two extreme cases using abundance matching
to CDM and to $\psiDM$ MFs, respectively. 

\begin{figure}[t!]
\centering
\includegraphics[width=8.5cm]{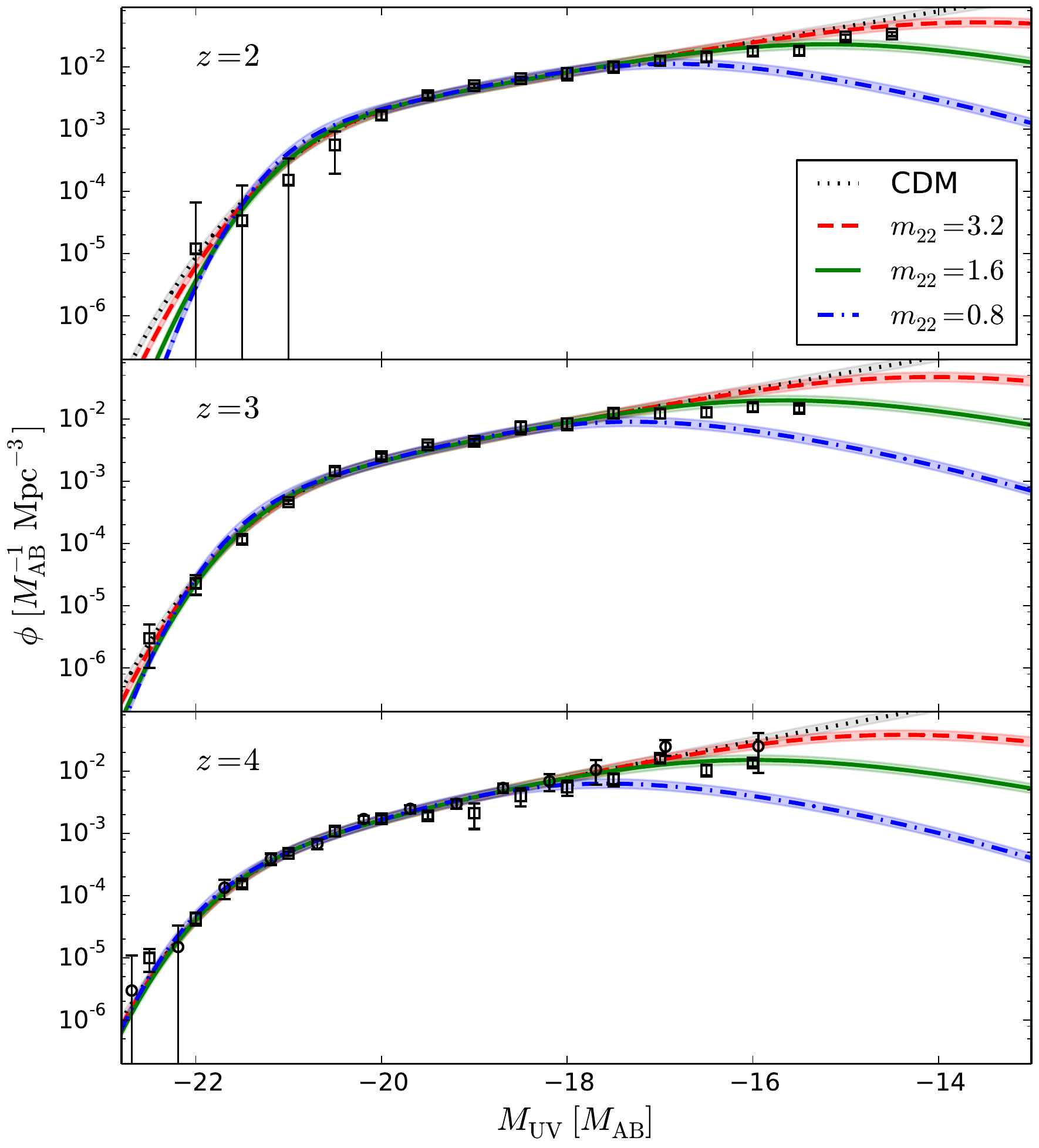}
\caption{
Luminosity function (LF) at $z=2-4$ predicted by the conditional LF model.
The shaded regions indicate the $2\sigma$ uncertainties. The data with $2\sigma$ error bars
are the LFs determined from B15b (open circles) and \citet[][open squares]{Parsa2015}.
The observed faint-end slope in this redshift range is consistent with the
$\psiDM$ model with $\PM \sim 1.6$ (see text). Notice that CDM overestimates
the number density of faint galaxies, especially at $z=3$.
}
\label{fig:LF2to4}
\end{figure}

It is also interesting to extend the comparison of LF to $z<5$
(Fig. \ref{fig:LF2to4}). To be self-consistent, we adopt the same
parameters of the conditional LF model shown in Table 1.
\citet{Parsa2015} recently found a much shallower faint-end slope at
$z=2-4$, distinctly different from the steep slope reported previously
\citep{Reddy2009,Alavi2014,Bouwens2015b}. $\psiDM$ with
$\PM \sim 1.6$ is found to provide a clear better fit to this shallower slope, while CDM
overestimates the number density of faint galaxies, especially at $z=3$. It is thus
very important for future research to understand this apparent discrepancy
between the faint-end slopes determined by \citet{Parsa2015} and
the previous studies at $z<5$.

It should be emphasized that the conditional LF model provides a more
reasonable constraint on the $\psiDM$ particle mass. At first glance,
it may seem that this approach has too many free parameters to provide an
appropriate estimation. However, actually the only relevant parameter for
constraining $\PM$ is $p$, the faint-end slope of the \ML relation, as argued
below. Since the faint-end slope of the LF is insensitive to $\Sigma$, we have
$L \sim L_c \propto \Mvir^{p}$. This leads to
$\phi(\MUV)=0.4\ln(10)L\phi(L)=0.4\ln(10)p^{-1}\dndlnM$,
where $\dndlnM$ is the halo MF in logarithmic mass bins.
Therefore, for a given $p$, a maximum observed $\phi(\MUV)$ can be directly
converted to a minimum required peak $\dndlnM$, which then turns into
a lower limit of $\PM$. Moreover, if $\dndlnM \propto \Mvir^\eta$, which is
appropriate when $\Mvir \gg M_0$ in \eref{eq:MFFit} (i.e., when $\psiDM$
is still close to CDM), then $\phi(L) \propto L^{\eta/p-1}$. Since $\eta$
can be estimated by the ST99 MF, we can determine
$p$ from a given $\phi(L)$. Accordingly, \emph{$p$ itself is not unconstrained.}

As an example, consider the $\psiDM$ model with $\PM=1.6$ at $z=6$.
The observed LF has $\phi(L) \propto L^{-2.1}$ at $\MUV \sim -19$.
This luminosity corresponds to a halo mass of $\Mvir \sim 7\times10^{10} \Msun$,
at which $\dndlnM \propto \Mvir^{-1.5}$ (see Fig. \ref{fig:MF}). We thus
have $p=-1.5/(-2.1+1) \sim 1.4$, consistent with Table 1. The MF
has a peak of $\dndlnM \sim (1.0 \pm 0.3)\times10^{-2} \Mpc^{-3}$ around
$\Mvir = 10^{10} \Msun$, which can then be converted to a maximum LF of
$\phi(\MUV) = (6.8 \pm 2.0)\times 10^{-3} \Mpc^{-3}$.
This estimation is in excellent agreement with \fref{fig:CondLF},
even though the only assumption made here is that the \ML relation
is a power law at the faint end.
Therefore, we conclude that the conditional LF model provides
a more plausible and less model-dependent estimation of the
$\psiDM$ particle mass than with abundance matching.

\subsubsection{Truncated Schechter Function}
\label{subsubsec:ModScheFunc}

It is also useful and convenient to parameterize the predicted LF of $\psiDM$
by a formula similar to the Schechter function. We adopt
\be
\phi(L) = \frac{\pstar}{\Lstar} \left( \frac{L}{\Lstar} \right)^\alpha
             \exp\left( -\frac{L}{\Lstar} \right) \Gamma(L),
\label{eq:ModScheFunc_Main}
\ee
where
\be
\Gamma(L) = \left[ 1 + \left( \frac{L}{\Lpsi} \right)^{\gamma} \right]^{\beta/\gamma}
\label{eq:ModScheFunc_Supp}
\ee
represents the suppression of faint galaxies in $\psiDM$ ($\Gamma=1$ for CDM).
$\Lpsi$ is the characteristic luminosity of the suppression, below which
$\phi(L)$ asymptotes to $L^{\alpha+\beta}$. To describe the
time evolution of LF, we follow the literature
\citep{Bouwens2012,Kuhlen2012,Schultz2014,Bouwens2015a} and assume that the parameters
in Equations (\ref{eq:ModScheFunc_Main}) and (\ref{eq:ModScheFunc_Supp}) depend
linearly on redshift.
Applying chi-square fitting to the conditional LF model
(Fig. \ref{fig:CondLF}) then leads to
\begin{eqnarray}
\Mstar &=& -20.90 - 0.004(z-6) \nonumber \\
\pstar &=& 0.52\times10^{-0.28(z-6)-3} \Mpc^{-3} \nonumber \\
\alpha &=& -1.78 - 0.06(z-6) \nonumber \\
\Mpsi  &=& -17.44 + 5.19\log(\PM/0.8) - 2.71\log((1+z)/7) \nonumber \\
\beta  &=& 1.69 + 0.03(z-6) \nonumber \\
\gamma &=& -1.10,
\label{eq:ModScheFunc_Para}
\end{eqnarray}
where we have assumed $\MUV = -2.5 \log(L/\rm erg \, s^{-1} \, Hz^{-1})+51.6$.

\begin{figure}[t!]
\centering
\includegraphics[width=8.5cm]{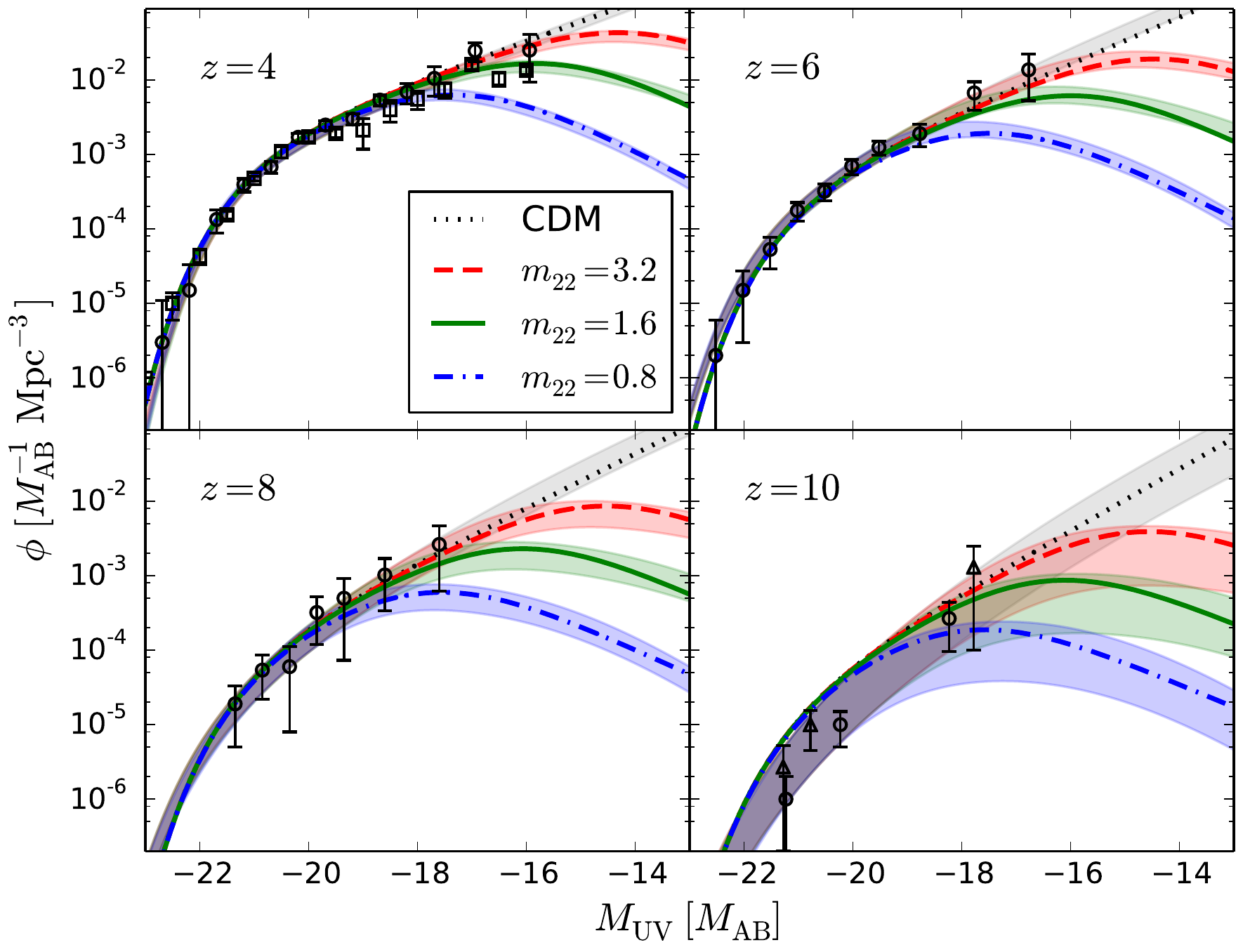}
\caption{Luminosity function (LF) at $z=4-10$ obtained by a single
analytic formula similar to the Schechter function
(Eqs. [\ref{eq:ModScheFunc_Main}-\ref{eq:ModScheFunc_Para}]; central lines).
The shaded regions are the same as Fig. \ref{fig:CondLF}, showing the LF
predicted by the conditional LF model within $2\sigma$.
Error bars show the observed LFs ($2\sigma$ at $z=4-8$ and $1\sigma$ at $z=10$)
of \citet[][open squares]{Parsa2015}, B15b (open circles), and \citet[][open triangles]{Oesch2014}.
The analytic formula well reproduces the conditional LF results at $z=4-8$,
while at $z=10$ it slightly outnumbers the observed galaxies and is
marginally consistent with the conditional LF model.
}
\label{fig:ModScheFunc}
\end{figure}

\fref{fig:ModScheFunc} shows the LF obtained by the truncated Schechter
function described above. It fits well with the LF predicted by the
conditional LF model at $z=4-8$, while for the bright end at $z=10$ it slightly outnumbers
the observed galaxies (see B15b and references therein) and
is marginally consistent with the conditional LF model.
This subtle discrepancy mainly results from the assumption of linear
dependence on redshift in \eref{eq:ModScheFunc_Para}, and it is unclear whether
it indicates a faster evolution at $z > 8$ given the substantial
uncertainties in the LF at $z=10$.
This possible acceleration in evolution
has been successfully modeled recently in the context of abundance matching
for CDM by additionally incorporating early stellar evolution \citep{Mason2015}.
It may be interesting to apply this approach in the context of $\psiDM$ too for extending
predictions to $z \ge 10$ with some security.

\subsection{Magnification Bias for the Hubble Frontier Fields}
\label{subsec:MagBias}

We have seen above that the quantum pressure inherent to $\psiDM$ leads to a
suppression of low-mass galaxies and hence our predictions for the LF have
largest contrast with CDM at low luminosities. Here we examine the benefits of
gravitational magnification in the new HFF data, where statistical samples of
multiply lensed galaxies are magnified by typically $\sim 10$ \citep{Lam2014},
reaching two magnitudes or more further down the LF at high redshifts. It
corresponds to an intrinsic UV luminosity of $M_{UV} \sim -15$, where we
predict sizeable difference between CDM and $\psiDM$ for the interesting
range of $\PM$ needed to provide the kpc-scale dark cores of local dSph
galaxies.

Gravitational lensing induces a bias in the number density of sources
detected above a flux limit, which in the case of lensing clusters is well
established \citep{Broadhurst1995,Umetsu2008} and has led to the detection
of the highest redshift and lowest luminosity galaxies currently known
\citep{Zheng2012,Coe2013,Zitrin2014}. The number density of such high-redshift
galaxies is modified in the following way:
\be
N_{\rm lensed}(\mathord{>}L) = {(1/{\mu})}N_{\rm unlensed}(\mathord{>}L/{\mu}),
\label{eq:MagBias}
\ee
where $\mu$ is the magnification factor and $N(\mathord{>}L)$ is the galaxy number density
above a flux limit corresponding to $L$. It shows the competition between the
enhanced number density due to the lower, magnified limiting luminosity and
the diminished source plane area, which is smaller than the observed area
by the same magnification factor. The magnification bias can thus be
defined as $N_{\rm lensed}(\mathord{>}L) / N_{\rm unlensed}(\mathord{>}L)$, which equals one
when there is no bias.

\begin{figure}[t!]
\centering
\includegraphics[width=8.5cm]{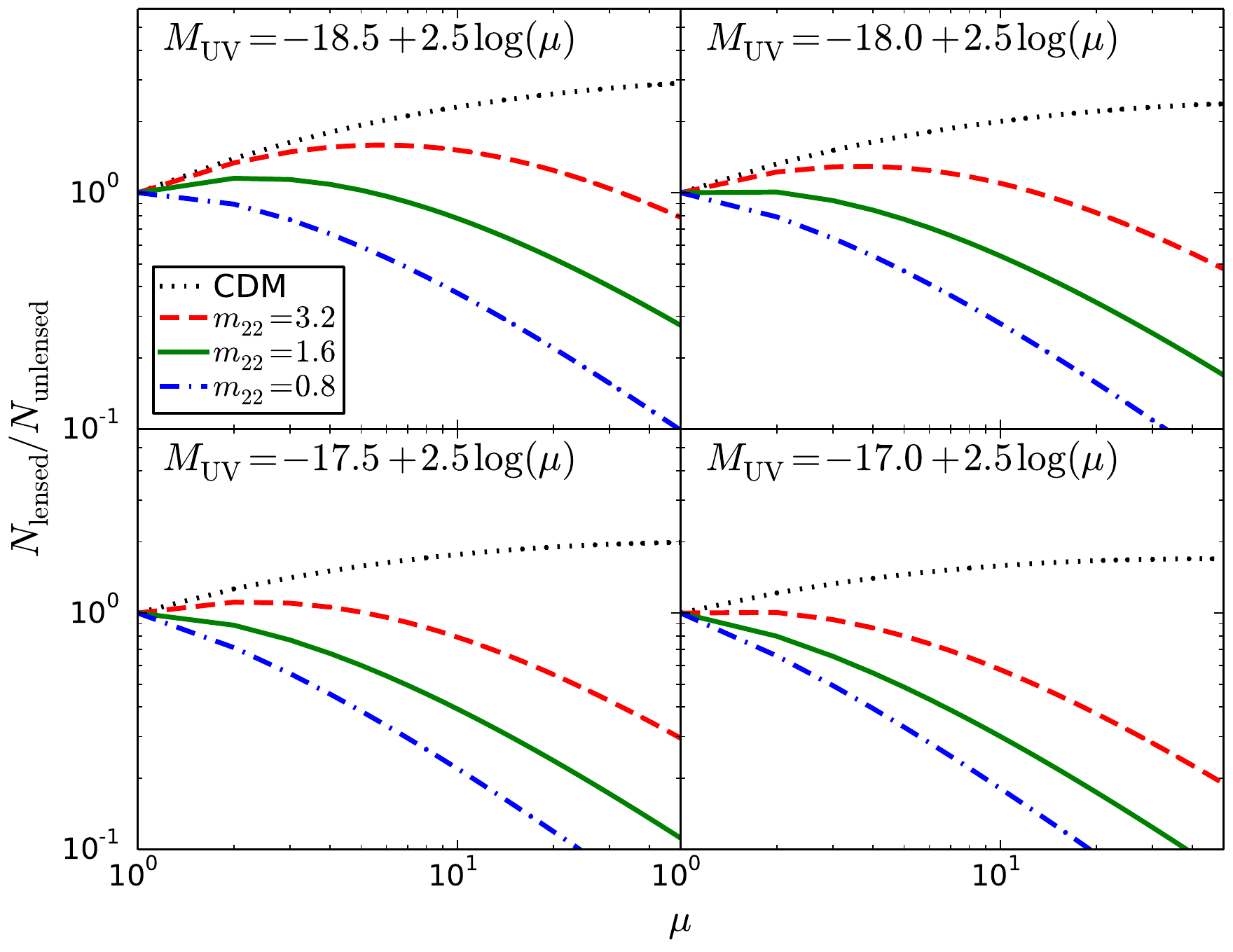}
\caption{
Magnification bias predicted by $\psiDM$ and CDM at $z=10$. The numbers on
the top of each panel indicate the adopted limiting luminosities. For CDM the
magnification bias continually rises owing to the steep faint-end slope of the
LF, whereas for $\psiDM$ the bias is generally lower than one because of the
strong suppression of low-mass halos. The lower the limiting luminosity, the
greater the contrast between $\psiDM$ and CDM.
}
\label{fig:MagBias}
\end{figure}

\fref{fig:MagBias} shows the difference between the magnification bias for
CDM and $\psiDM$ at $z=10$, which we predict to be a strong function of limiting
luminosity because of the difference in sign in the faint-end slope of the LFs
between these two models. CDM is continually rising and hence the
magnification bias is greater than one, thereby enhancing the number of faint
galaxies detected at high redshifts, whereas for $\psiDM$ the turnover in the
LF leads to many fewer galaxies magnified above the flux limit.
The difference we predict at $\MUV=-14.5$ for example is a factor of
$\sim 10$ for $\mu \sim 10$ at $z \sim 10$ between CDM and $\psiDM$ with
$\PM \sim 1.2$. Going beyond this with JWST should probe another two
magnitudes deeper with the assistance of lensing \citep{Mason2015} - most
efficiently by employing the same deep lenses as the HFF for
which the magnification maps have been widely studied \citep{Rodney2015} and
best understood \citep{Lam2014,Diego2015}.

\subsection{Reionization}
\label{subsec:Reionization}

Based on the predicted $\psiDM$ LF, we
can calculate the reionization history using the standard approach that has been adopted for
various dark matter models \citep[e.g.,][]{Kuhlen2012,Schultz2014,Bozek2015}. The time
evolution of the volume filling fraction of ionized hydrogen, $\QHII(z)$,
is governed by
\be
\frac{d\QHII}{dt} = \frac{\nidot}{\nHbar} - \frac{\QHII}{\trec},
\label{eq:QHII}
\ee
where $\nHbar$ is the mean comoving hydrogen number density. We take
$\QHII=0$ at $z=25$. The volume-averaged recombination time, $\trec(z)$,
can be determined by
\be
\trec \sim 0.93\left( \frac{\CHII}{3} \right)^{-1}
               \left( \frac{1+z}{7} \right)^{-3} {\rm Gyr},
\label{eq:trec}
\ee
where $\CHII \equiv \langle \nHII^2 \rangle / \langle \nHII \rangle^2$ is the
volume-averaged clumping factor. Here we have assumed an intergalactic medium
temperature of $2 \times 10^4\,{\rm K}$ and a primordial hydrogen mass
fraction of $0.76$.

The comoving ionizing emissivity, $\nidot(z)$, defined as the number of ionizing
photons produced per unit time per unit comoving volume, can be estimated from
the galaxy UV LF:
\be
\nidot = \frac{2 \times 10^{25}}{\rm erg\,Hz^{-1}} \, \zetai \fesc \int_{\Llim}^{\infty}
         dL \phi(L) L,
\label{eq:niondot}
\ee
where $\zetai$ represents the efficiency of converting galaxy UV luminosity to
ionizing photon luminosity, and $\fesc$ is the escape fraction. Strictly speaking,
since the observed rest-frame UV luminosity (at $\sim$ 1500\,\AA) will also be 
extinguished by dust, $\fesc$ in \eref{eq:niondot} is the \emph{relative} escape
fraction \citep{Steidel2001} defined as the \emph{absolute} escape fraction 
(the fraction of `ionizing photons' that escapes the galaxy without being absorbed
by dust and neutral hydrogen) divided by the fraction of `UV photons' that
escapes. This relative escape fraction can be significantly higher than the absolute
escape fraction because of the efficient dust extinction at $\sim$ 1500\,\AA.
$\Mlim = -2.5 \log(\Llim/\rm erg \, s^{-1} \, Hz^{-1})+51.6$ is the limiting
UV magnitude, below which the galaxy formation is assumed
to be inefficient. Note that $\nidot$ is sensitive to the faint-end
slope of LF that differentiates various dark matter models.

For a given $\QHII(z)$, the Thomson optical depth to CMB can be calculated via
\be
\taue = c \, \sigma_{\rm T} \, \nHbar \int_0^\infty dz \frac{(1+z)^2}{H(z)} \QHII(z)
        (1 + \eta(z) Y/4X),
\label{eq:taue}
\ee
where $\sigma_{\rm T}$ is the Thomson cross-section, $H(z)$ is the Hubble parameter,
$X \sim 0.76$ and $Y=1-X$ are the primordial mass fraction of hydrogen and
helium, respectively, and we take $\eta(z>4)=1$ when helium is only singly ionized
and $\eta(z\le4)=2$ when helium is doubly ionized by quasars.

There are three free parameters in Equations (\ref{eq:QHII})-(\ref{eq:taue}), namely,
$\CHII$, $\Mlim$, $\zetai \fesc$ ($\zetai$ and $\fesc$ are fully degenerate), which,
for simplicity, are assumed to be spatially uniform and redshift-independent in this
work. The typical parameter ranges adopted in the literature are
$\CHII=2\sim5$, $\Mlim=-17\sim-10$, $\zetai=0.5\sim2.0$, and $\fesc=0.1\sim0.5$
\citep{Bouwens2012,Kuhlen2012,Schultz2014,Bozek2015,Bouwens2015a,Robertson2015}.
To bracket the
uncertainties in these parameters, we consider three different parameter sets: a minimum
reionization model (MIN) with $(\CHII,\zetai \fesc,\Mlim)=(4.0,0.2,-13)$, a fiducial
reionization model (FID) with $(\CHII,\zetai \fesc,\Mlim)=(3.0,0.6,-13)$, and a maximum 
reionization model (MAX) with $(\CHII,\zetai \fesc,\Mlim)=(2.0,1.0,-13)$. The adopted
values are also shown in Table 2. Note that since $\psiDM$ should be
insensitive to $\Mlim$ (because of the strong suppression of small halos),
we fix $\Mlim=-13$ unless otherwise specified.

\begin{table}
\label{tab:reionpara}
\begin{center}
\caption{Reionization parameters}
\begin{tabular}{lccc}
\hline
Model & $\CHII$ & $\zetai \fesc$ & $\Mlim$\tablenotemark{a} \\
\hline
MIN & 4.0 & 0.2 & -13 \\
FID & 3.0 & 0.6 & -13 \\
MAX & 2.0 & 1.0 & -13 \\
\hline
\end{tabular}
\end{center}
\tablenotemark{a} $\Mlim$ is allowed to vary when computing $\taue$.
\end{table}

\begin{figure}[t!]
\centering
\includegraphics[width=8.5cm]{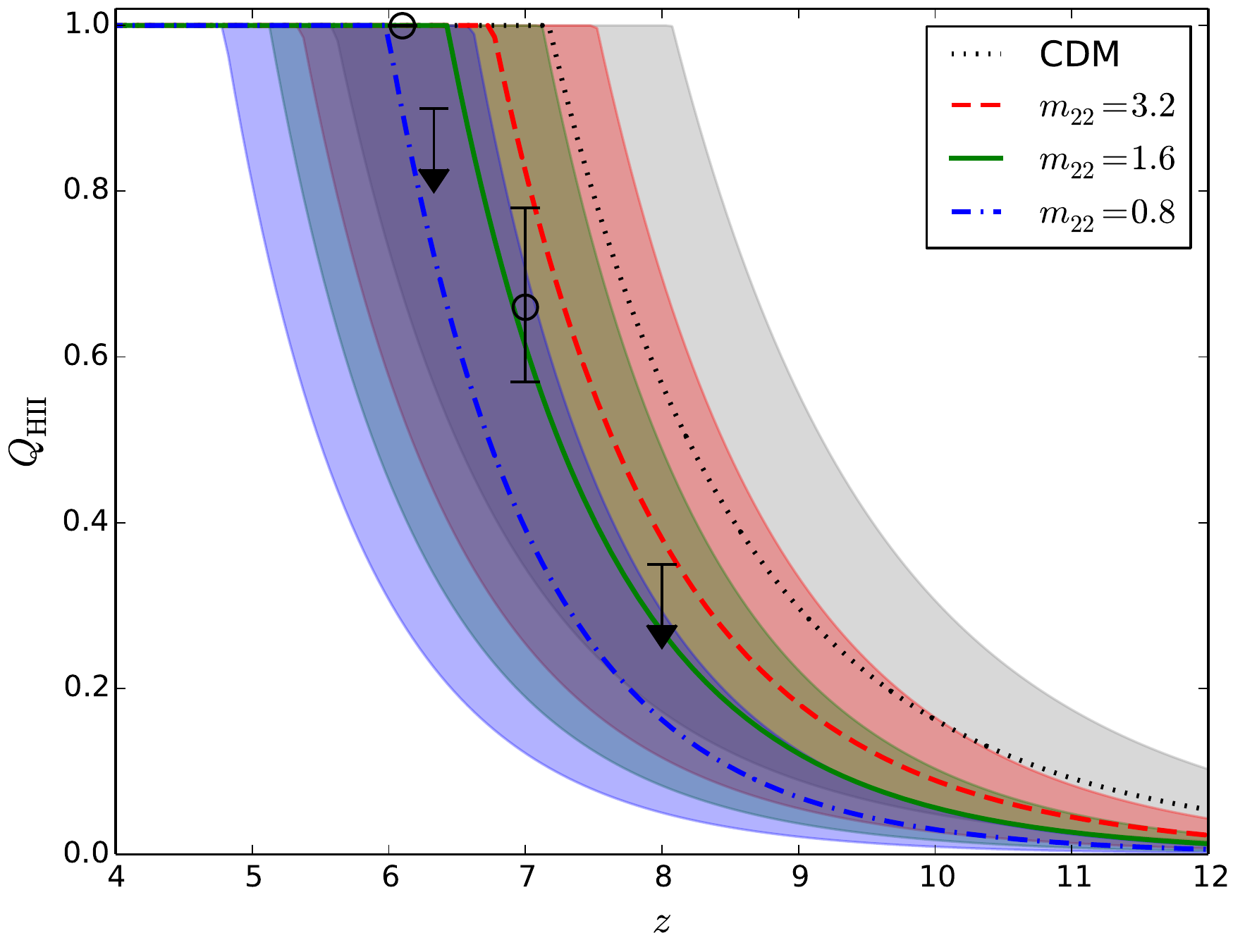}
\caption{Volume filling fraction of HII as a function of redshift. The central lines 
correspond to the FID model and the shaded regions are bounded by the
MIN and MAX models. The limiting UV magnitude is fixed to $\Mlim=-13$.
Open circles and arrows mark the observational constraints
at $z \sim 6-8$ \citep{Fan2006,Schroeder2013,Schenker2014,McGreer2015}.
}
\label{fig:QHII}
\end{figure}

\fref{fig:QHII} shows the time evolution of $\QHII$ for various models.
For a given reionization parameters, $\psiDM$ predicts a faster increase
of ionized volume at later times. Recent observations indicate that reionization
undergoes a rapid evolution at $6<z<8$ and is completed at $z \sim 6$
\citep{Fan2006,Schroeder2013,Schenker2014,McGreer2015}. Correspondingly,
the required ionizing photon production efficiency, $\zetai \fesc$, is
$\sim 0.6$ for $\psiDM$ with $\PM=0.8$,
about two times higher than CDM
($\zetai \fesc \sim 0.3$).
For $\psiDM$ this result is insensitive to the values of
both $\Mlim$ and $\CHII$ probed in this work.

\begin{figure}[t!]
\centering
\includegraphics[width=8.5cm]{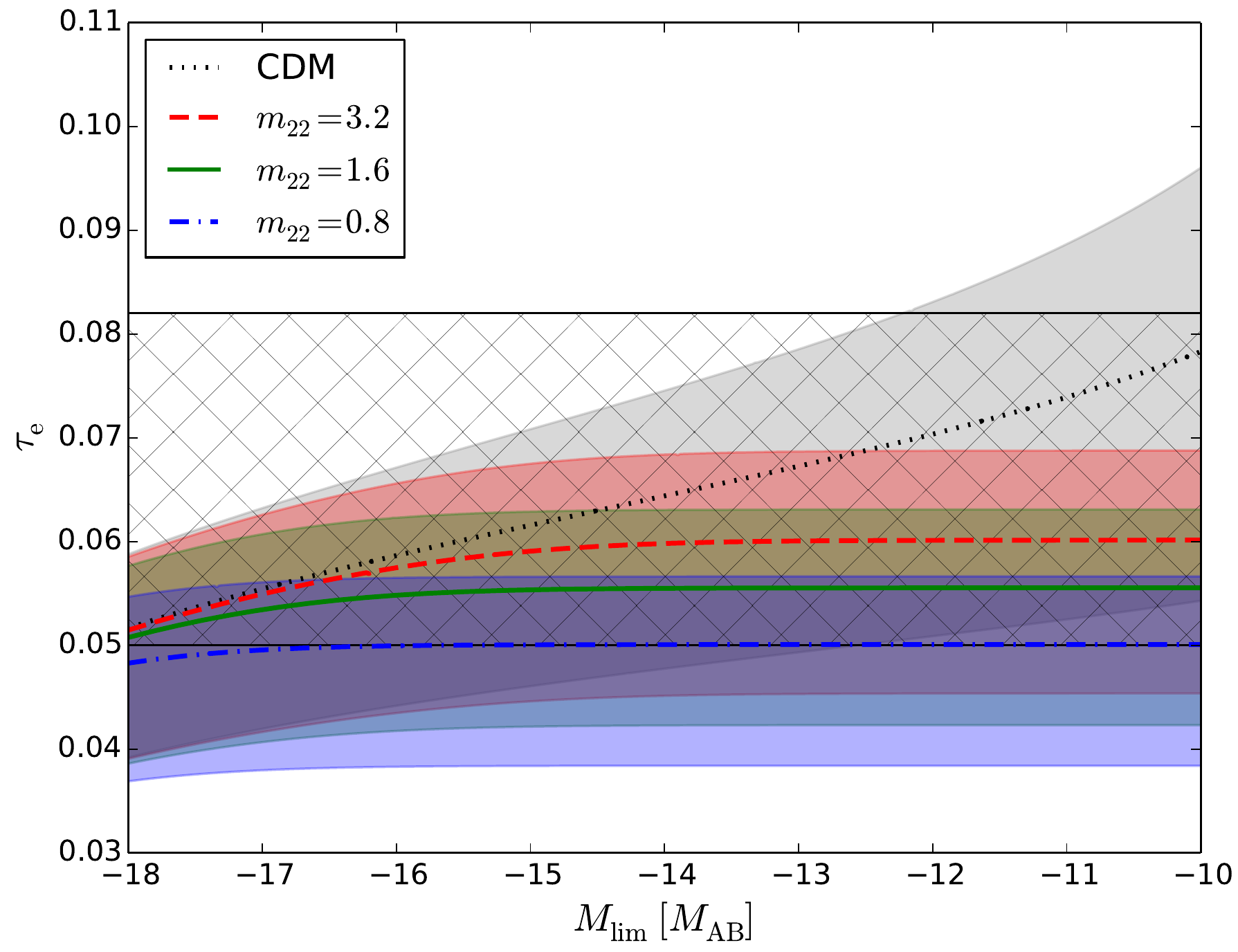}
\caption{Thomson optical depth versus limiting UV magnitude. The central lines
correspond to the FID model and the shaded regions are bounded by the
MIN and MAX models, where $\Mlim$ is allowed to vary. The cross-hatched
region shows the Planck 2015 $1\sigma$ confidence limit \citep{Planck2015}.
}
\label{fig:taue}
\end{figure}

\fref{fig:taue} shows the Thomson optical depth as a function of $\Mlim$ for
various models. Most recent Planck 2015 results give
$\taue = 0.066 \pm 0.016$ \citep[$1\sigma$,][]{Planck2015},
which is marginally consistent with $\PM=0.8$ assuming
$\zetai \fesc \sim 0.6$, in agreement with the estimation from $\QHII(z)$.
Note that this estimation is largely independent of both $\CHII$ and $\Mlim$
when $\Mlim \gtrsim -15$, since galaxies fainter than this magnitude
are highly suppressed at $z\ge6$ even for $\PM = 3.2$.
Given the considerable uncertainties in the reionization model at high
redshifts, it is therefore clear that neither $\QHII(z)$ nor $\taue$
provides a stringent constraint on $\psiDM$ even for $\PM \sim 0.8$.
In comparison, CDM with an ionizing photon production efficiency as low as
$\zetai \fesc \sim 0.2$ can be consistent with Planck 2015 assuming
$\Mlim \gtrsim -13$, in agreement with the findings of
\citet{Bouwens2015a} and \citet{Robertson2015}.

\citet{Bozek2015} used a similar approach to estimate the $\psiDM$ particle mass.
They reported that $\PM = 1.0$ is disfavored by the observed value of $\taue$ 
at $3\sigma$, and consequently a particle mass as high as $\PM = 10$ is required,
significantly higher than our estimation. This inconsistency with our result mainly arises
from a higher value of $\taue$ they adopted from the previous Planck 2013 results,
$\taue = 0.090 \pm 0.013$ \citep[$1\sigma$,][]{Planck2013,Spergel2015}.
Also note that the LF we predict based on the `conditional LF model'
declines at the faint end (see Fig. \ref{fig:CondLF}),
while the LF adopted by \citet{Bozek2015} based on the `abundance matching'
approach does not have
this natural feature. Such relative deficit of faint galaxies in our LF model
would only delay the reionization process and reduce the Thomson optical depth,
hence increasing the discrepancy between observations and the $\psiDM$ model
with a small particle mass. However, here we demonstrate that $\PM \sim 0.8$
can still be consistent with the latest Planck observations provided that the ionizing
photon production efficiency is sufficiently high.

\section{DISCUSSION AND CONCLUSION}
\label{sec:Discussion}

In this paper, we have constructed cosmological simulations designed to study the dark matter
halo MF in the wave dark matter ($\psiDM$) scenario. Here
the uncertainty principle counters gravity below a Jeans scale, which is
determined by the only free parameter in this model, $\PM$, the dark matter
particle mass. The smaller the particle mass, the larger the Jeans
scale, and hence the stronger the suppression of low-mass halos.
For this reason, we focus on determining a lower limit of $\PM$ based
on the observed UV LF at $z \sim 4-10$ and the reionization history.

The major findings in this study can be briefly summarized as follows:
\begin{itemize}
\item{$\psiDM$ halo MF has a prominent drop below $\sim 10^{10} \Msun$,
which can be well fitted by \eref{eq:MFFit}.}

\item{$\psiDM$ predicts a clear drop in the galaxy LF around $\MUV \gtrsim -16$
at $z \gtrsim 4$ based on a conditional LF model, which can be fitted by
a truncated Schechter function
(Eqs. [\ref{eq:ModScheFunc_Main}-\ref{eq:ModScheFunc_Para}]).}

\item{The newly established LF at $z\sim4-10$ constrains the $\psiDM$
particle mass to be $\PM \ge 1.2$ ($2\sigma$).}

\item{For galaxies magnified $\mathord{>}10\times$ in the Hubble Frontier Fields, $\psiDM$
predicts an order of magnitude fewer detections than CDM at $z \gtrsim 10$
down to an intrinsic UV luminosity of $\MUV \sim -15$}.

\item{$\psiDM$ with $\PM \gtrsim 0.74$ can satisfy the Thomson optical depth
reported by the latest Planck observations, on the assumption of a
reasonable ionizing photon production rate.}
\end{itemize}

In the following we give a more thorough discussion of this work.
We first argue that, for studying the $\psiDM$ MF with the
particle masses, redshift range, and halo masses of interest here
($\PM \sim 1$, $z\sim 4-10$, $\Mvir \gtrsim 1\times10^9 \Msun$),
it is reasonable to approximate both the $\psiDM$ transfer function as
redshift-independent and the evolution of quantum fluid by collisionless
particles. The major drawback of these approximations is the inability
to capture the difference in the substructures between CDM and $\psiDM$
halos, where $\psiDM$ halos have prominent solitonic cores in the centers
surrounded by fine-scale, large-amplitude cellular interference (SCB14a, S14b).
However, this shortcoming is irrelevant when
one is only concerned with the halo masses, as in this work. This is
especially true because most halos above $\sim 10^9 \Msun$ have yet to
merge gravitationally with each other at $z \gtrsim 4$.

Simulations of collisionless particles with a cut-off in the initial power
spectrum suffer from the well-known side effect of inducing spurious halos due to
numerical artefacts, which must be accounted for when determining an accurate MF
at low masses. To identify and remove these artificial halos,
we adopt a similar approach suggested by \citet{Lovell2014} based
on the shape of the progenitors of halos and the spatial overlap between
low-resolution halos and their high-resolution counterparts (see Appendix \ref{sec:SpuriousHalo}).
The resulting MF cleaned in this way then shows a clear decline
as expected below the Jeans mass and can be well fitted by \eref{eq:MFFit}.
Most importantly, this reinforces the MF we derive above
$\gtrsim 3\times10^9 \Msun$, which are most relevant for comparing with
observations.

Comparing the halo MF with the observed UV LF
requires the knowledge of \ML relation. For this purpose, we first apply the
abundance matching technique to either CDM or $\psiDM$ MFs. In
both cases, $\psiDM$ with $\PM=0.8$ shows a deficit of low-mass galaxies with
$\Mvir \sim 10^{10} \Msun$ at $z=6-8$ (see Fig. \ref{fig:CumuMF}), leading to
lower limits of particle mass, $\PM \ge 1.5$ (abundance matching to CDM) and
$\PM \ge 0.9$ (abundance matching to $\psiDM$). We also explore the conditional
LF model as an alternative approach, which yields $\PM \ge 1.2$. The key
assumption here is a power-law \ML relation at the faint end.
We argue that this approach provides a more reasonable and
model-independent estimation of the $\psiDM$ particle mass. In addition, we
predict that the high-z LF should turn over slowly around $\MUV \gtrsim -16$
at $z \gtrsim 4$, distinctly different from CDM.
This predicted feature lies just beyond the detected luminosity range of the
current LFs at $z \gtrsim 4$, but will be directly testable
with forthcoming observations such as JWST and also with highly magnified
galaxies in the HFF data.

Note that the \ML relation for low-mass halos may be subject to large
uncertainties. \citet{Strigari2008} showed that the Milky Way dwarf
satellites, which have $\Mvir \lesssim 10^{10} \Msun$,
share a common mass scale but have luminosity differences over four orders of
magnitude. Therefore, a more complicated \ML relation might be expected for
halos below $\sim 10^{10} \Msun$, at least at lower redshifts.
Even in CDM simulations, \citet{OShea2015} found a relatively flat high-z
LF at the faint end compared with the Schechter function fits to observations
(B15b). Ideally, $\psiDM$ simulations with the addition of baryonic
physics may be very helpful in further differentiating the high-z LFs predicted
by CDM and $\psiDM$.

The Thomson scattering optical depth to CMB provides another constraint for
the $\psiDM$ particles mass. In general, $\psiDM$ predicts a faster increase
of ionized volume at later times due to the suppression of early galaxy
formation. We demonstrate that $\psiDM$ with $\PM \gtrsim 0.74$ can satisfy
the Planck 2015 results and have reionization completed at $z \gtrsim 6$, on
the assumption that the ionizing photon production rate is sufficiently
efficient (about three times higher than that required for CDM).
This result is largely independent of the limiting luminosity and the faint-end
slope adopted since galaxies fainter than $\MUV \sim -15$ are highly suppressed
at $z\ge6$ even for $\PM = 3.2$.
On the other hand, this constraint is somewhat undermined because of
the current large uncertainties associated with the escape fraction
and the efficiency of converting galaxy UV luminosity to ionizing photon
luminosity. Note that for CDM the reionization calculation is much more
uncertain in fact than for $\psiDM$, as it is dominated by the choice of
limiting luminosity assumed for the relatively steeply rising
LF ($\alpha >1.0$) as otherwise the integrated luminosity diverges.

The MF below $\sim 10^9 \Msun$ determined by our simulations has a larger
relative uncertainty in our model. Firstly, approximating the $\psiDM$
transfer function as redshift-independent can somewhat underestimate the
matter power spectra at higher redshifts (see Fig. \ref{fig:EvolvePS}),
which in turn leads to a underestimation of MF at low masses.
Secondly, we approximate the evolution of quantum fluid by collisionless
particles. It may allow for the formation of a small number of halos with
masses well below the Jeans mass, which would otherwise be suppressed further
if the dynamical effect of quantum pressure is taken into account.
In this sense, the MF in a bona-fide wave-based $\psiDM$ simulation may have
an even stronger break at the low-mass end than \fref{fig:MF}.
We may hope to check on this in the future by running the
full adaptive-mesh-refinement wave simulations that we have previously described
(SCB14a, S14b) on a substantially more powerful platform.
Thirdly, there are also uncertainties associated with the
removal of spurious halos below $\sim 10^9 \Msun$ (see the shaded areas in
Fig. \ref{fig:MF}). However, it should be emphasized that none of these
uncertainties are relevant for the purpose of this study at the level of
precision that is currently afforded by the data, as the depths of the
current Hubble and forthcoming JWST do not extend to such low-mass halos.

Though not entirely consistent, the $\psiDM$ particle mass estimated in
this work, $\PM \ge 1.2$ ($2\sigma$), is surprisingly close to the values
determined from local dwarf galaxies.
SCB14a established with the first wave-based $\psiDM$
simulations a distinct solitonic core in the center of every
$\psiDM$ halo. We have previously obtained $\PM=0.8\pm0.2$ ($1\sigma$) by
fitting the spatial distribution of the intermediate metallicity stellar
population in the Fornax dSph galaxy to the soliton mass profile, under
the assumption of a constant projected velocity dispersion. \citet{Marsh2015a}
have also determined a similar constraint, $\PM \le 1.1$ ($2\sigma$), by
fitting the mass profiles of Fornax and Sculptor dSph galaxies to the soliton
mass profile using an empirical relation between the half-light radius and
velocity dispersion. Note that for $\PM=1.2$, we predict that a halo with
$\Mvir = 2\times10^9 \Msun$ still has a core as large as $\sim 1.1 \kpc$ (S14b),
consistent with many estimates of the large cores found
in dSph galaxies \citep[e.g.,][]{Salucci2012,Amorisco2013}.
A more thorough comparison between the stellar phase-space distribution in
dSph galaxies and the $\psiDM$ halo mass profile using a full Jeans analysis
will be extremely important to further clarify how coincident are the
particle masses determined by these various approaches and the role of
baryonic feedback in this context (S-R. Chen et al., in preparation).
The inherent density granularity of the $\psiDM$ halo may also be examined
through internal dynamical effects and by lensing flux anomalies on sub-kpc
scales, which provide other key independent observational tests for
distinguishing $\psiDM$ from WDM and CDM.

The finding that a similar particle mass in $\psiDM$ can both solve the
cusp-core problem in dwarf galaxies \citep{Moore1994} and satisfy the high-z
LF and reionization observations is very encouraging for this model.
It demonstrates a great advantage of $\psiDM$ over WDM, for which the light
particle mass required for creating kpc-scale cores in dSph galaxies prevents
the formation of the host dwarf galaxies in the first place and overly
suppresses high-z galaxies \citep{Maccio2012,Schneider2014}.
The key reason for this striking difference is that the relation between
core radius and power spectrum truncation differs because of the very
different physics underlying $\psiDM$ and WDM, with the uncertainty principle
responsible in the case of $\psiDM$ and free streaming in the case of WDM.
The particle masses in the two models can always be chosen so that they have
similar truncated matter power spectra, but the corresponding core radius in
$\psiDM$ can then be several times larger than that in WDM. This is why
\emph{$\psiDM$ does not suffer from the Catch 22 problem affecting WDM}.

\section{ACKNOWLEDGEMENT}
We thank David Marsh for stimulating discussions.
This work is supported in part by the National Science Council of Taiwan
under the grant MOST 103-2112-M-002-020-MY3.

\appendix
\section{REMOVAL OF SPURIOUS HALOS}
\label{sec:SpuriousHalo}

Particle simulations with an initial power spectrum cut-off suffer from the
formation of spurious halos, especially at low masses. It is therefore
necessary to adopt a robust algorithm to identify genuine halos in the
simulations, which we describe below.

\begin{figure}[t!]
\centering
\includegraphics[width=8.5cm]{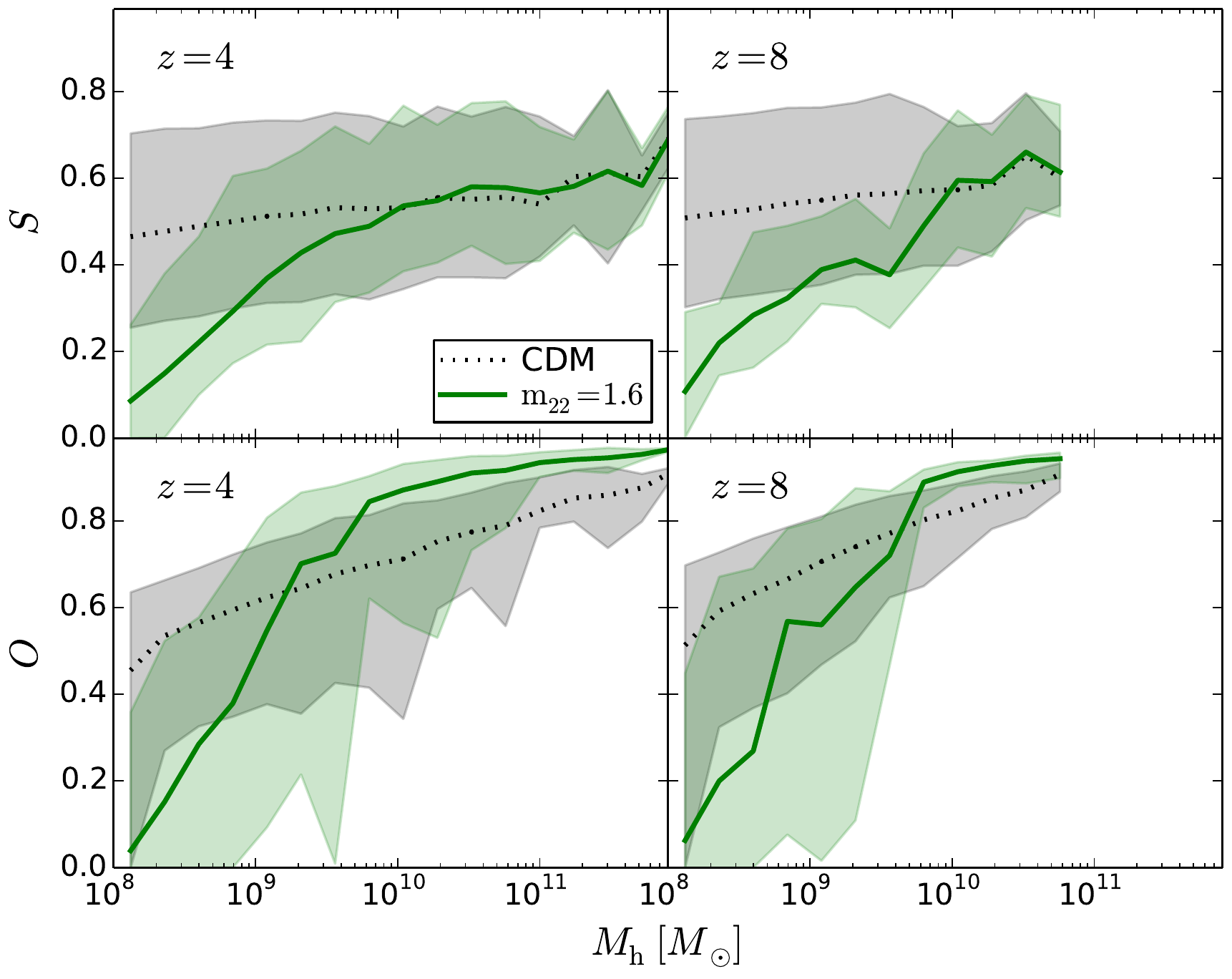}
\caption{
Sphericity and spatial overlap factor as a function of halo mass in the
$15 \Mpch$ simulations. The central solid and dotted lines show the mean
values, and the shaded areas contain $95\%$ of halos. Low-mass $\psiDM$
halos have lower sphericity and overlap factor compared with CDM halos,
especially for $\Mvir \lesssim 10^9 \Msun$, indicative of a severe
contamination from spurious halos in this mass range.
}
\label{fig:SO_Level}
\end{figure}

We first use the AMIGA Halo Finder \citep[AHF,][]{Knollmann2009} to
identify all halos (both genuine and spurious halos), and find the
corresponding protohalos by tracing the constituting particles back to the
initial redshift. We then define the protohalo sphericity as
\be
S = \sqrt{\frac{I_1+I_2-I_3}{-I_1+I_2+I_3}},
\label{eq:Sphericity}
\ee
where $I_1 \le I_2 \le I_3$ are the principle moments of inertia of a
protohalo. This is equivalent to having $S=c/a$, where $a$ and $c$ are
the maximum and minimum semi-axis lengths of an ellipsoid of uniform density
with the same principle moments of inertia as the protohalo
\citep{Lovell2014}. \fref{fig:SO_Level} (upper panel) shows the sphericity
as a function of halo mass for both CDM and $\psiDM$ with $\PM=1.6$ in the
$15\Mpch$ simulations with $512^3$ particles. Clearly, CDM halos have
higher $S$. The average $S$ of $\psiDM$ halos lies below the
$2\sigma$ lower limit of CDM halos for $\Mvir \lesssim 10^9 \Msun$, 
indicative of a severe contamination from spurious $\psiDM$ halos in
this mass range.

To quantify the matching accuracy between two protohalos in low- and
high-resolution simulations, respectively, we introduce a spatial
overlap factor,
\be
O = \frac{ V_{\rm low} \cap V_{\rm high} } { V_{\rm low} \cup V_{\rm high} },
\label{eq:Overlap}
\ee
where $V$ is the volume of a protohalo estimated by depositing particles onto
grids using the cloud-in-cell scheme with a cloud size equal to the mean
particle separation. $V_{\rm low} \cap V_{\rm high}$ and $V_{\rm low} \cup V_{\rm high}$
represent the intersection and union of two volumes, respectively.
Accordingly, $O=1$ if the spatial distribution of two protohalos are
completely overlapped, while $O=0$ if there is no overlap at all.
Note that our definition of spatial overlap factor differs from that of
\citet[][see Eq. 6 therein]{Lovell2014}. \eref{eq:Overlap} does not require
solving the gravitational potential and thus is more computationally
efficient and more sensitive to the actual spatial overlap between two
protohalos. \fref{fig:SO_Level} (lower panel) shows the spatial overlap factor
as a function of halo mass. The average $O$ of $\psiDM$ halos shows an
apparent drop at $\Mvir \lesssim 10^9 \Msun$, again suggesting that
spurious halos start to outnumber genuine halos.

We consider most CDM halos as genuine since its MF is consistent with the
analytic prediction of ST99 above $\sim 3\times10^8 \Msun$
(corresponding to $\sim 100$ particles in our lower resolution
simulations). We thus define the $2\sigma$ lower
limit of $S$ and $O$ at $\Mvir = 3\times10^8 \Msun$ in CDM as the minimum
thresholds, $\Scut$ and $\Ocut$, below which halos are regarded as spurious.
$\Scut$ and $\Ocut$ slightly increase with redshift, and we take
$\Scut = 0.31$ and $\Ocut = 0.34$ by averaging over $z=4-10$.
Note that we do not use the mass cut criterion adopted by \citet{Lovell2014},
which relies on extrapolating the low-resolution results to high-resolution
runs.

In short, for each low-resolution halo identified by AHF, we calculate
$S$ by \eref{eq:Sphericity} and determine the
highest value of $O$ by applying \eref{eq:Overlap} for all nearby
high-resolution halos. Then, we use CDM halos to determine the minimum
thresholds, $\Scut$ and $\Ocut$, and only low-resolution halos with
$S \ge \Scut$ and $O \ge \Ocut$ are marked as genuine. \fref{fig:HaloMap}
(lower panel) shows the spatial distribution of genuine halos.
Clearly, suspicious low-mass halos, most visible along filaments, are
removed with the above procedure, while only massive halos with a good match
between low- and high-resolution simulations are retained.

\bibliographystyle{apj}
\bibliography{Reference}

\end{document}